\documentclass[floats,floatfix,showpacs,amssymb,prd,twocolumn,superscriptaddress,nofootinbib
]{revtex4-2}

\usepackage{graphicx,epsf, epsfig, amssymb}
\usepackage{bm}
\usepackage{longtable,tabularx}
\usepackage{color}
\definecolor{linkcolor}{rgb}{0.0,0.3,0.5}
\usepackage[unicode, colorlinks=true, linkcolor=linkcolor, citecolor=linkcolor, filecolor=linkcolor,urlcolor=linkcolor, pdfusetitle, bookmarksopen=true]{hyperref}
\usepackage{amsfonts,amsmath,amssymb,mathrsfs,gensymb}
\usepackage{natbib}
\usepackage{prd_macros}
\usepackage{array}
\usepackage{multirow}
\usepackage{rotating,array}
\usepackage{graphicx}
\usepackage[normalem]{ulem}
\usepackage[caption=false]{subfig} 
\usepackage{ulem}
\usepackage{dcolumn}
\usepackage{braket}
\usepackage{appendix}
\usepackage{comment}
\usepackage{siunitx}

\usepackage [english]{babel}
\usepackage [autostyle, english = american]{csquotes}
\MakeOuterQuote{"}

\usepackage{tabularx}

\def\apj{\rm{Astrophys.~J.}}
\def\apjl{\rm{Astrophys.~J.~ Lett.}}
\def\apjs{\rm{ApJS}}
\def\aap{\rm{Astron.~Astrophys.}}

\def\mnras{\rm{Mon.~Not.~R.~Astron.~Soc.}}
\def\prd{\rm{Phys.~Rev.~D}}
\def\prl{\rm{Phys.~Rev.~Lett.}}

\def\pasp{\rm{PASP}}

\def\nat{\rm{Nature}}

\newcommand\orcid[1]{\href{https://orcid.org/#1}{$\!$\includegraphics[scale=0.013]{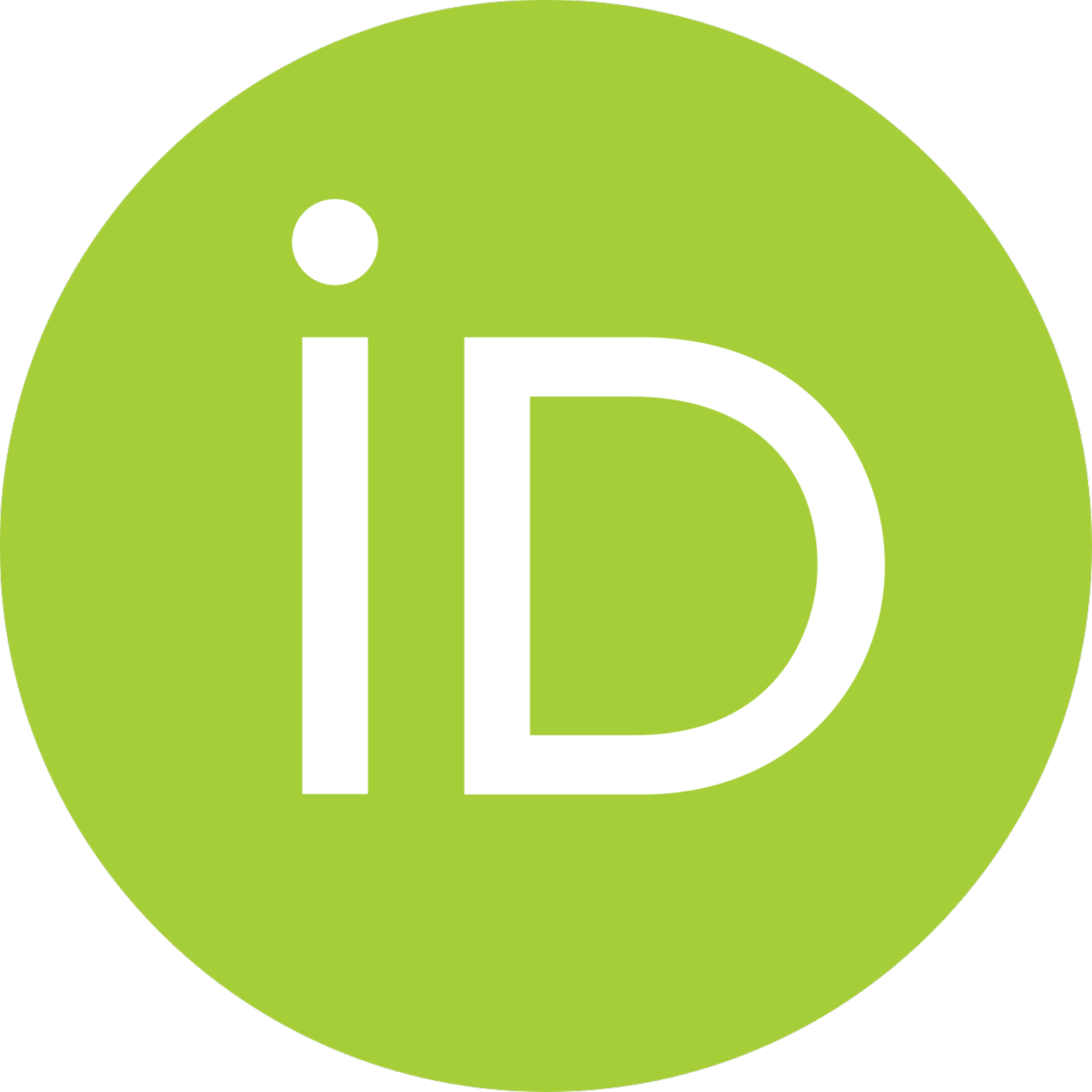} $\!\!$}}

\newcommand{\bham}{\affiliation{School of Physics and Astronomy \& Institute for Gravitational Wave Astronomy, \\University of Birmingham, Birmingham, B15 2TT, UK}}
\newcommand{\dallas}{\affiliation{Department of Physics, The University of Texas at Dallas, Richardson, Texas 75080, USA}}

\begin{document}

\setlength{\parskip}{0.2pt}

\title{Signatures of spin precession and nutation in isolated black-hole binaries} 

\author{Nathan Steinle \orcid{0000-0003-0658-402X}}
\email{nsteinle@star.sr.bham.ac.uk}
\bham
\dallas

\author{Michael Kesden \orcid{0000-0002-5987-1471}}
\email{kesden@utdallas.edu}
\dallas

\date{\today}

\begin{abstract}
The spin precession of binary black holes (BBHs) that originate from isolated high-mass binary stars is determined by the interplay of phenomena such as tides, winds, accretion, common-envelope evolution, natal kicks, and stellar core-envelope coupling. In previous work, we identified regions of the parameter space that may produce BBHs with large misalignments from natal kicks and high spin magnitudes from three mechanisms - tides, accretion, or inheritance via minimal core-envelope coupling.  Here, we explore the spin precession of such BBHs using five parameters that describe the amplitude and frequency with which the orbital angular momentum precesses and nutates about the total angular momentum, modulating the gravitational-wave emission. Precession is generally possible for sufficiently strong natal kicks provided at least one of the black holes is spinning. Nutation is a consequence of spin-spin coupling and depends on the three spin-up mechanisms. Tidal synchronization can leave a distinct correlation between the aligned effective spin and the nutation frequency, but does not produce large nutations. When a black hole accretes $\gtrsim 20\%$ of its companion's envelope, the precession frequency and amplitude are large. A much smaller amount of accretion, e.g., $\approx 2\%$, is needed to provide a large precession frequency and amplitude when the accretor is a Wolf-Rayet (WR) star. The inheritance of high natal WR spins ($\gtrsim 5\%$ of their maximum breakup value) via minimal core-envelope coupling is the most promising mechanism for producing nutating BBHs, implying that a measurement of nutation from gravitational-wave observations may suggest isolated-binary origin with minimal core-envelope coupling.
\end{abstract}

\maketitle

\section{Introduction} \label{sec:Intro}

Studies of the formation and evolution of black holes have been bolstered by the detection of gravitational waves (GWs) that are emitted during the coalescence of stellar-mass binary black holes (BBHs) by the LIGO/Virgo detectors. Two main channels of formation for BBHs are expected theoretically \cite{Benacquista2013,Postnov2014}. If the individual black holes originate before the binary is formed via dynamical interactions in a dense stellar cluster, then the directions of the BBH spins are expected to be isotropic. This implies that the spin precession of the BBH is "generic", i.e. both the precession and nutation amplitudes of the orbital angular momentum can be large as it precesses about the total angular momentum. On the other hand, if the BBH forms as the product of isolated stellar binary evolution, then the spin orientations are traditionally thought to be at least partially aligned with the binary orbital angular momentum \cite{Rodriguez2016b,Gerosa2018}, implying that the orbital angular momentum might have modest precession and nutation amplitudes. However, in \citeauthor{Steinle2021}~\cite{Steinle2021}, we explored four pathways of isolated stellar binary evolution and identified regions of the parameter space from which BBHs may emerge with high dimensionless spin magnitudes, either from tidal synchronization, accretion, or inheritance, and with large spin-orbit misalignments from natal kicks. This suggests that measurement of large, misaligned BBH spins in a GW event may not be a ``smoking gun'' indicating a dynamical origin. Distinguishing between the likely formation channels will require at least hundreds of BBH detections \cite{Zevin2017,Bouffanais2019,Mapelli2020}, and many frameworks exist to constrain the fractional contributions of each channel \cite{Breivik2016,Talbot2017,Stevenson2017,Vitale2017,Zevin2017,Taylor2018,Bouffanais2019,Mandel2019,Zevin2020b,Wong2021,Mould2022}.

A clear identification of spin precession has been reported in only one LIGO/Virgo event \cite{LIGO2021catalog,Hannam2021,Varma2022b} thus far, GW200129, and recent studies indicate that BH spins are likely not small in the existing dataset, e.g.'s, \cite{LIGO2021astro,Callister2022,Mould2022b}. Spin precession has been thoroughly studied, typically with timescale hierarchies in the post-Newtonian regime \cite{Thorne1980,Cutler1994,Apostolatos1994,Damour2001,Kesden2015}. The work of \citeauthor{Kesden2015}~\cite{Kesden2015} and \citeauthor{Gerosa2015b}~\cite{Gerosa2015b} demonstrated that the orbital angular momentum nutates about the total angular momentum when the total spin magnitude $S$ is time dependent, and that the amplitude of this nutation is correlated with the morphology of the spin components in the orbital plane. Their work was recently extended by \textcolor{linkcolor}{Gangardt~\&~Steinle $et~al.$}~\cite{GangardtSteinle2021}, which presented five intuitively and geometrically derived spin-precession parameters that are constant on the precession timescale and encapsulate the precession and nutation of the binary orbital angular momentum. These five precession parameters were shown to robustly represent spin-precession phenomenology.

The purpose of this work is to quantify the precession and nutation of BBHs that originate from the isolated stellar binary evolutionary pathways of \citeauthor{Steinle2021}~\cite{Steinle2021} with the five spin precession parameters of \textcolor{linkcolor}{Gangardt~\&~Steinle $et~al.$}~\cite{GangardtSteinle2021}. We accomplish this by demonstrating the dependence of the five precession parameters - the precession amplitude $\langle \theta_L \rangle$\,, the precession frequency $\langle\Omega_L\rangle$\,, the nutation amplitude $\Delta\theta_L$\,, the nutation frequency $\omega$\,, and the precession-frequency variation $\Delta\Omega_L$\, - on the relevant initial stellar binary parameters - the natal kick strength $\sigma,\,$ the initial separation $a_{\rm ZAMS},\,$ the accreted fraction $f_{\rm a},\,$ and the breakup-spin fraction $f_{\rm B}$ - of our astrophysical model of BBH formation. We find that precession is possible given sufficiently strong natal kicks, depending on the pathway of evolution. Highly precessing systems can emerge from accretion in stable mass transfer, but tides are not efficient at producing precessing or nutating systems. Nutation is generally possible when the stellar progenitors evolve under inefficient stellar angular momentum transport. This suggests that nutating binaries can generically originate from the isolated formation channel.

This paper is organized as follows: in section \ref{sec:Meth}, we review the model of isolated BBH formation (subsection \ref{subsec:AstroPaper}) and the model of BBH spin precession (subsection \ref{subsec:SpinPre}); in section \ref{sec:Results}, we discuss the dependence of the BBH spin precession parameters and morphologies on the various initial stellar binary parameters, we present a few illuminating examples; and we conclude with a summary and discussion of implications for GW observations and predictions of BBH formation in section \ref{sec:Discussion}.

\section{Methodology} \label{sec:Meth}

\subsection{Isolated black-hole binary formation}
\label{subsec:AstroPaper}

For a detailed description of our model of isolated BBH formation, see our previous work \cite{Steinle2021}, from which we only briefly state the main assumptions and findings here. A zero-age main-sequence (ZAMS) binary star is initialized at the binary separation $a_{\rm ZAMS}$ with metallicity $Z$, and with masses $m_{1, \rm ZAMS}$ of the primary star and $m_{2, \rm ZAMS}$ of the secondary star.  We define $m_{1, \rm ZAMS} \geq m_{2, \rm ZAMS}$ such that the ZAMS mass ratio is $q_{\rm ZAMS} = m_{2, \rm ZAMS}/m_{1, \rm ZAMS} \leq 1$. 

Various astrophysical effects and processes are parameterized to identify regions of the isolated stellar binary initial parameter space that lead to precessing BBHs. We assume either $maximal$ (strong) stellar core-envelope spin coupling, resulting in low natal BH spins, or $minimal$ (weak) coupling in which the BH natal spin is set by the initial spin of the newly born Wolf-Rayet (WR) star parameterized by the fraction $f_{\rm B}$ of its breakup value. We assume that Roche lobe overflow (RLOF) immediately initiates a mass-transfer event either through common-envelope evolution (CEE), which drastically shrinks the binary separation, or in stable mass transfer (SMT), where a fraction $f_{\rm a}$ of the donor's envelope is accreted by the companion. We assume that supernova (SN) natal kicks, whose kick velocity magnitude depends on the 1-dimensional Maxwellian dispersion parameter $\sigma$, introduce spin-orbit misalignments that are essential for producing precessing systems in our model. For convenience, we refer to the process of stellar collapse and BH formation generically as SN regardless of whether they are accompanied by a luminous transient. We enforce the Kerr limit on the dimensionless spin of the collapsing progenitor, by assuming that angular momentum is lost either with (isotropic) or without (negligible) accompanying mass loss.

\begin{figure}[!t]
\centering
\includegraphics[width=0.48\textwidth]{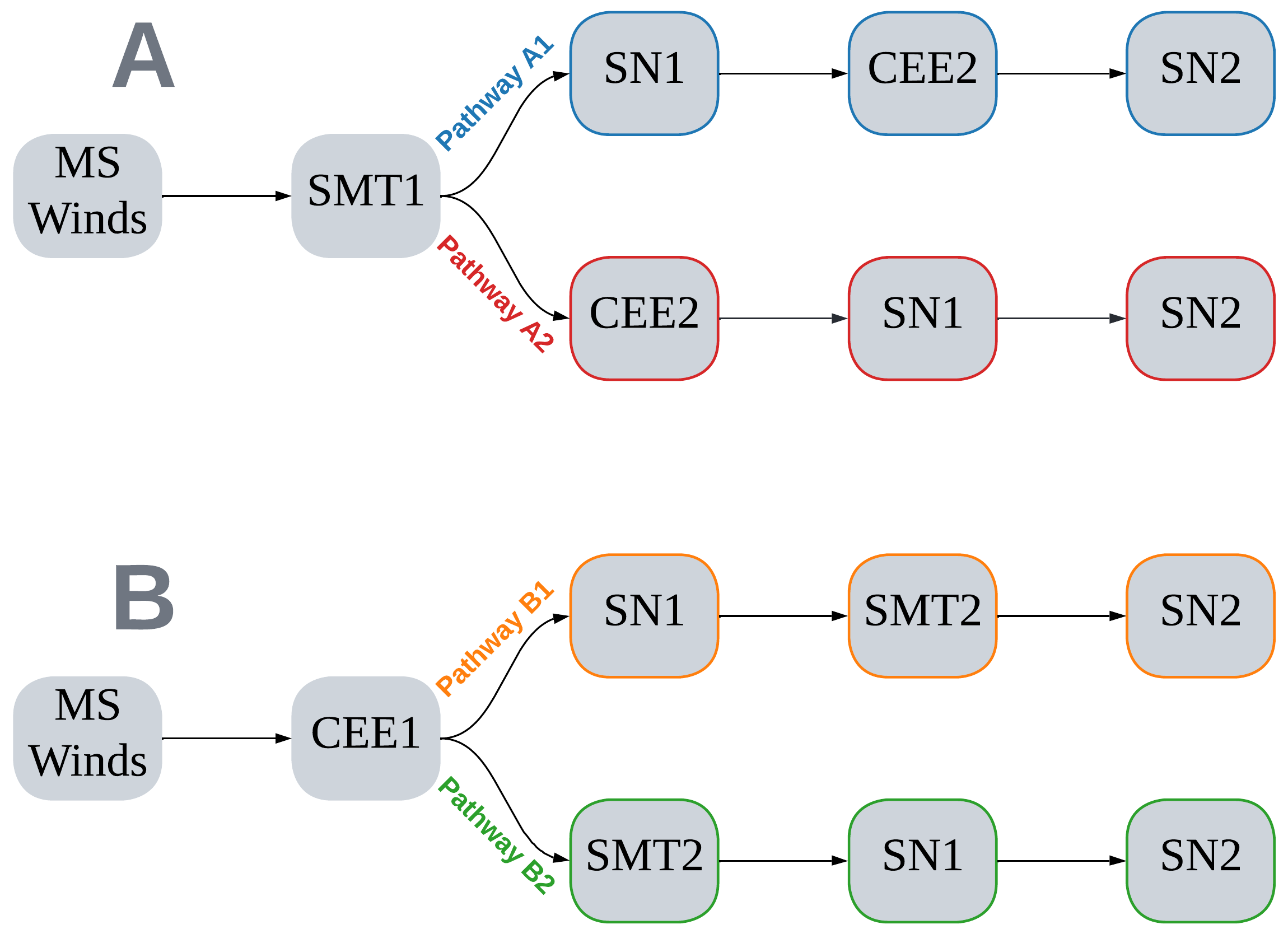}
\caption{
Diagram of the four evolutionary pathways that we explore. $\it{Top}$: Evolutionary scenario A, where the primary RLOF leads to stable mass transfer (SMT1), and the secondary RLOF leads to common-envelope evolution (CEE2). $\it{Bottom}$: Evolutionary scenario B, where the primary RLOF leads to common-envelope evolution (CEE1), and the secondary RLOF leads to stable mass transfer (SMT2). In Pathway 1 (Pathway 2) of each scenario, the primary supernova SN1 occurs before (after) secondary RLOF.
} \label{F:Diagram}%
\end{figure}

We explore two scenarios of isolated binary evolution defined by whether RLOF of the primary leads to SMT (Scenario A) or to CEE (Scenario B). As depicted in Figure~\ref{F:Diagram}, both scenarios allow for two unique pathways of binary stellar evolution depending on whether the supernova of the primary (SN1) occurs before (Pathway 1) or after (Pathway 2) RLOF of the secondary (CEE2 in Scenario A, SMT2 in Scenario B). Pathway 1 occurs in binaries with ZAMS mass ratio below a transition value $q_{\rm trans}$, where as always in our work we define mass ratios to be less than unity. Pathways A2, B1, and B2, in which CEE precedes the first natal kick, dominate at large values of $\sigma$ because kicks can more readily unbind binaries in Pathway A1 where SN1 occurs before CEE has shrunk the orbit. We indicate the order of evolutionary events for each pathway, as shown in Fig.~\ref{F:Diagram}, in the figure captions of Section~\ref{sec:Results}, e.g., Pathway A1 corresponds to SMT1-SN1-CEE2-SN2. Mass-ratio reversal (MRR) occurs when the primary star evolves into the less massive BH, e.g.'s, \cite{Zevin2022,Broekgaarden2022}. In Pathways A1 and A2, it depends crucially on the fraction $f_{\rm a}$ of the donor's envelope that is accreted during stable mass transfer, while in Pathway B2 it also depends on Kerr-limit mass loss in BH formation of the primary star (i.e., isotropic mass loss causes MRR, and negligible mass loss does not).

Spin-orbit misalignments are necessary for BBH spin precession, while nutation also requires that both BHs have high spin magnitudes since it is a consequence of spin-spin coupling \cite{Apostolatos1994}. In our model, BHs can acquire significant misalignments from SN natal kicks whose magnitude is parameterized by $\sigma$ and whose direction is isotropic. BHs can acquire high spin magnitudes through three mechanisms: $tides$, where the WR progenitor is spun up by tides exerted by its companion at the small binary separations that follow CEE; $accretion$, where the BH or the WR progenitor gains angular momentum by accreting from its companion during SMT; and $inheritance$, where, assuming weak core-envelope coupling, the BH inherits the rotational angular momentum of its WR progenitor which is parameterized by the fraction $f_{\rm B}$ of its break-up value. Inheritance can produce high natal BH spins in either scenario if weak core-envelope coupling is assumed \cite{Maeder2004,Maeder2008,Maeder2012,Belczynski2020}, tides can only produce high WR spins in Scenario A, since the small separations where the tidal torque is strong are disallowed in Scenario B due to the early onset of CEE, and accretion can only produce high spin in Scenario B, since spin gained from accretion by the secondary main sequence star in Scenario A is lost under strong core-envelope coupling or not inherited under weak core-envelope coupling. In our model, spin misalignments are uncorrelated with the breakup fraction $f_{\rm B}$ and are uncorrelated with the accreted fraction $f_{\rm a}$, although it has been shown that accretion can increase the donor's misalignment \cite{Stegmann2021}.

The effects of tides and accretion on the BBH spins are highly sensitive to the pathway of binary stellar evolution. In Pathway A1, only the secondary WR star can experience tides, since the primary BH forms before CEE occurs.  However, in Pathway A2, both WR components can be tidally synchronized and aligned for sufficiently small initial binary separation $a_{\rm ZAMS}$. In Pathway B1, the primary accretes as a BH and can obtain a high spin for adequately large accreted fraction $f_{\rm a}$. In Pathway B2, a much smaller value of $f_{\rm a}$ is sufficient to produce a maximally spinning primary BH, since the primary accretes as a WR star in this pathway which has much higher specific angular momentum at its surface than at the innermost stable circular orbit of a BH.

The Gaussian distribution of natal kicks introduces scatter in the initial BBH parameters, such that a single choice of parameters for the ZAMS binary yields a distribution of BBHs.
This scatter depends on the pathway of evolution, since the subsequent deterministic astrophysical processes are nonlinear functions of the post-kick parameters requiring systems to be tracked through BBH formation. In the results section of \cite{Steinle2021}, we only presented the average values of the BBH masses and spins at formation, but in this work we present the 5th, 50th, and 95th percentiles of the precession parameters at reference GW frequency $f = 20$~Hz.

After the natal kick of the secondary, we compute $t_{\rm merge}$ with the subsequent separation and eccentricity for each BBH, and we assume that the orbit is circularized due to GW emission \cite{Peters1964} to arrive at a distribution of BBHs with the parameters $(q,~M,~a_{\rm BBH},~\chi_{1},~\chi_{2},~\cos\theta_{1},~\cos\theta_{2})$. We use these quantities as inputs for the code \texttt{PRECESSION} \cite{Gerosa2016} that evolves each BBH through the PN inspiral, i.e., see Section~\ref{subsec:SpinPre}, starting from their initial separations $a_{\rm BBH} \gtrsim 10^4M$ and ending at a final mass- and frequency-dependent separation $r_f$. We assume a uniform distribution for the angle $\Delta\Phi$ that subtends the projections of the spin vectors in the orbital plane. Approximating the GW frequency, $f$, to be twice the orbital frequency, Kepler's Law provides $\pi Mf = \left(M/r_f\right)^{3/2}$ (with $G = c = 1$) and hence,
\begin{equation}\label{E:FinalSep}
\frac{r_f}{M} = 30\left(\frac{f}{20\,{\rm Hz}}\right)^{-2/3}\left(\frac{M}{{\rm 20\,M}_{\odot}}\right)^{-2/3}\,.
\end{equation}
If the total mass $M$ of a given BBH is too large, e.g., $M > M_{\rm max} = 102\,{\rm M}_\odot\left(20\,{\rm Hz}/f\right)$, then we do not use Eq.~(\ref{E:FinalSep}) since it would yield a separation that is smaller than the gravitational radius $r_{\rm g} = GM/c^2 \approx 10M$ where the PN approximation is no longer valid, and instead assume that ${r_f}/M = 10$. Throughout this work, a reference frequency of $f = 20$~Hz is assumed which gives a typical final separation $r_f \approx 15M$.

\subsection{Spin-precession formalism} \label{subsec:SpinPre}

BBHs inspiralling on quasi-circular orbits evolve on three distinct timescales: the binary separation $\mathbf{r}$ changes direction on the orbital timescale $t_{\rm orb}/M \sim (r/M)^{3/2}$, the two BBH spins $\mathbf{S}_{1}$ and $\mathbf{S}_{2}$ and the orbital angular momentum $\mathbf{L}$ precess on the precession timescale $t_{\rm pre}/M \sim (r/M)^{5/2}$, and the magnitude of binary separation changes on the radiation-reaction timescale $t_{\rm RR}/M \sim (r/M)^4$.  At lowest post-Newtonian (PN) order, spin precession modulates gravitational waveforms through the changing direction of $\mathbf{L}$.  As the direction of the total angular momentum $\mathbf{J} = \mathbf{L} + \mathbf{S}_{1} + \mathbf{S}_{2}$ remains nearly constant throughout the inspiral, except in the special case of transitional precession \cite{Apostolatos1994,Zhao2017}, the direction of $\mathbf{L}$ can be specified by the polar angle $\theta_L$ and azimuthal angle $\Phi_L$ in the frame of reference where $\mathbf{J}$ is along the z-direction.

The timescale hierarchy $t_{\rm orb} \ll t_{\rm pre} \ll t_{\rm RR}$ in the PN regime $r \gg M$ allows us to average all quantities evolving on the orbital timescale $t_{\rm orb}$ and hold constant all quantities evolving on the radiation-reaction timescale $t_{\rm RR}$ when calculating the evolution of $\theta_L$ and $\Phi_L$ on the intermediate precession timescale $t_{\rm pre}$ \cite{Kesden2015,Gerosa2015b}. On $t_{\rm pre}$, the polar angle $\theta_L(S)$, given by the law of cosines $\cos\theta_L(S) = (L + S_1\cos\theta_1 + S_2\cos\theta_2)/J = (J^2 + L^2 - S^2)/2JL$, is a function of the magnitude of the total spin $S = |\mathbf{S}_{1} + \mathbf{S}_{2}|$ which oscillates with nutation period $\tau = 2\int_{S_{-}}^{S_{+}}dS/|dS/dt|$, where the turning points $S_\pm$ depend on the BBH mass ratio $q$, spin magnitudes $S_1$ and $S_2$, $J = |\mathbf{J}|$, and the aligned effective spin \cite{Damour2001}
$\chi_{\rm eff} \equiv (\chi_{1}\cos\theta_{1} + q\chi_{2}\cos\theta_{2})/(1 + q)$, all of which remain constant on $t_{\rm pre}$. The rate $\Omega_L$ at which the azimuthal angle $\Phi_L$ changes is also a function of $S$ [Eq.~(29) of \cite{Gerosa2015b}]. Since the dynamics depend on only the single instrinsic variable $S$, any quantity that varies on $t_{\rm pre}$ may be averaged over the nutation period $\tau$ [Eq.~(33) of \cite{Gerosa2015b}], which we denote with angle brackets $\langle\cdot\rangle$. 

We calculate the following five parameters that encode the generic spin precession of a BBH: the precession amplitude $\langle \theta_L \rangle$\,, the precession frequency $\langle\Omega_L\rangle$\,, the nutation amplitude $\Delta\theta_L \equiv (\theta_{L+} - \theta_{L-})/2$\,, the nutation frequency $\omega \equiv 2\pi/\tau$\,, and the precession-frequency variation $\Delta\Omega_L \equiv (\Omega_{L+} - \Omega_{L-})/2$\,. These five spin precession parameters were presented in previous work and their behaviors explored for individual BBH inspirals and for stastiscally large distributions of BBHs with isotropically oriented spin directions \cite{GangardtSteinle2021} - as would be the case if they had originated from the dynamical formation channel. The behavior of the five spin precession parameters for our binary distributions is more easily understood by considering the values of $\cos\theta_1$ and $\cos\theta_2$ at $r/M \to \infty$, i.e., $\cos\theta_{1\infty}$ and $\cos\theta_{2\infty}$, to those of the binaries in \cite{GangardtSteinle2021} with similar spin magnitudes and mass ratio.

The precession (i.e., the azimuthal motion) and the nutation (i.e., the polar motion) of $\mathbf{L}$, and hence the five parameters, depend on the initial parameters of the BBH: large $\langle \theta_L \rangle$ occurs for BBHs with small mass ratio $q \leq 1$ and is larger for high dimensionless spin magnitudes $\chi_i$, large $\langle\Omega_L\rangle$ occurs for large $q$ and $\chi_i$, while $\Delta\theta_L$ is largest for $q \approx 0.6$ and $\chi_i \gtrsim 0.5$ and for spin orientations for which $\mathbf{J} \parallel \mathbf{L}$ at some point late in the inspiral which corresponds to the divergence of $\Delta\Omega_L$. The nutation parameters $\Delta\theta_L$ and $\Delta\Omega_L$ vanish as $q \to 0$ where the spin of the secondary becomes negligible, and as $q \to 1$ where $dS/dt$ vanishes. These possibilities are categorized in a ``taxonomy'' of spin precession where the presence of both precession and nutation of $\mathbf{L}$ is called ``generic'' precession, and the absence of nutation (i.e., $\Delta\theta_L = \Delta\Omega_L = 0$) is a special case called ``regular'' precession which includes single-spin systems, the equal-mass limit, and the spin-orbit resonances. Regarding observations of BBH sources, events that satisfy the above conditions may help to distinguish the signatures of precession and nutation in the waveform, since each provide corrections to the GW phase and each modulates the GW amplitude.

\section{Results}
\label{sec:Results}

\begin{figure*}[!t] 
  \centering
  \includegraphics[width=1.0\linewidth]{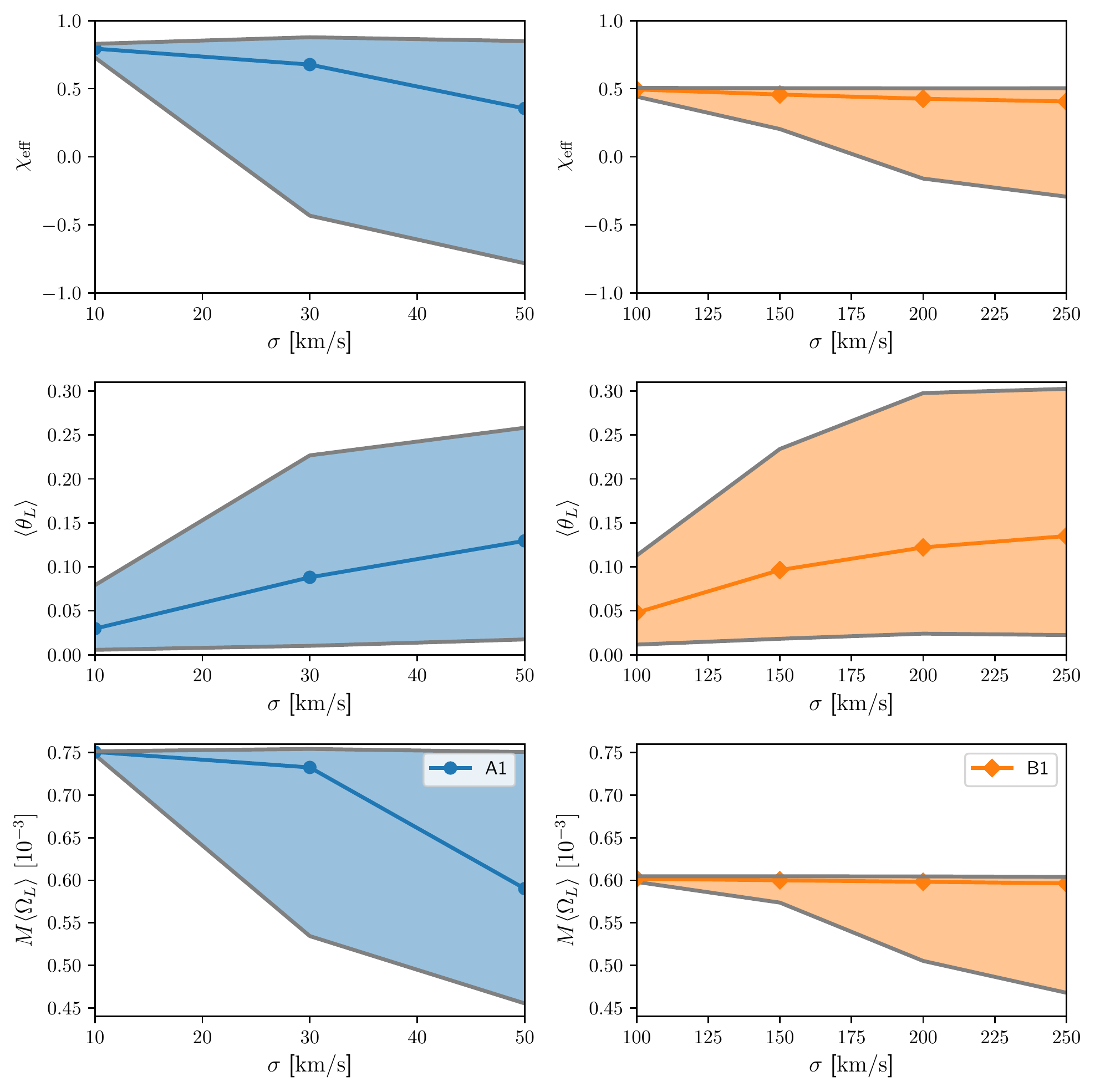}
   \caption{The dependence of the aligned effective spin $\chi_{\rm eff}$, average precession amplitude $\langle\theta_L\rangle$, and average precession frequency $\langle\Omega_L\rangle$ on the natal kick velocity dispersion $\sigma$, respectively, for BBHs that originate from Pathway A1 (B1), i.e., SMT1-SN1-CEE2-SN2 (CEE1-SN1-SMT2-SN2), assuming weak core-envelope coupling and negligible mass loss in BH formation due to the Kerr limit. The stellar binaries are initialized with $Z = 0.0002$, $m_{1,\rm ZAMS} = 70$~M$_{\odot}$, $m_{2,\rm ZAMS} = 50$~M$_{\odot}$, $f_{\rm a} = 0.2$, $a_{\rm ZAMS} = 6{,}000$ ($a_{\rm ZAMS} = 12{,}000$), and $f_{\rm B} = 0.05$ ($f_{\rm B} = 0.01$). The blue (orange) region indicates 90\% of BBHs in each distribution that evolved from Pathway A1 (B1), and the solid colored line marked by circles (diamonds) is the median percentile.
   } \label{F:VsSigma}
\end{figure*}

\begin{figure*}[!t] 
  \centering
  \includegraphics[width=1.0\linewidth]{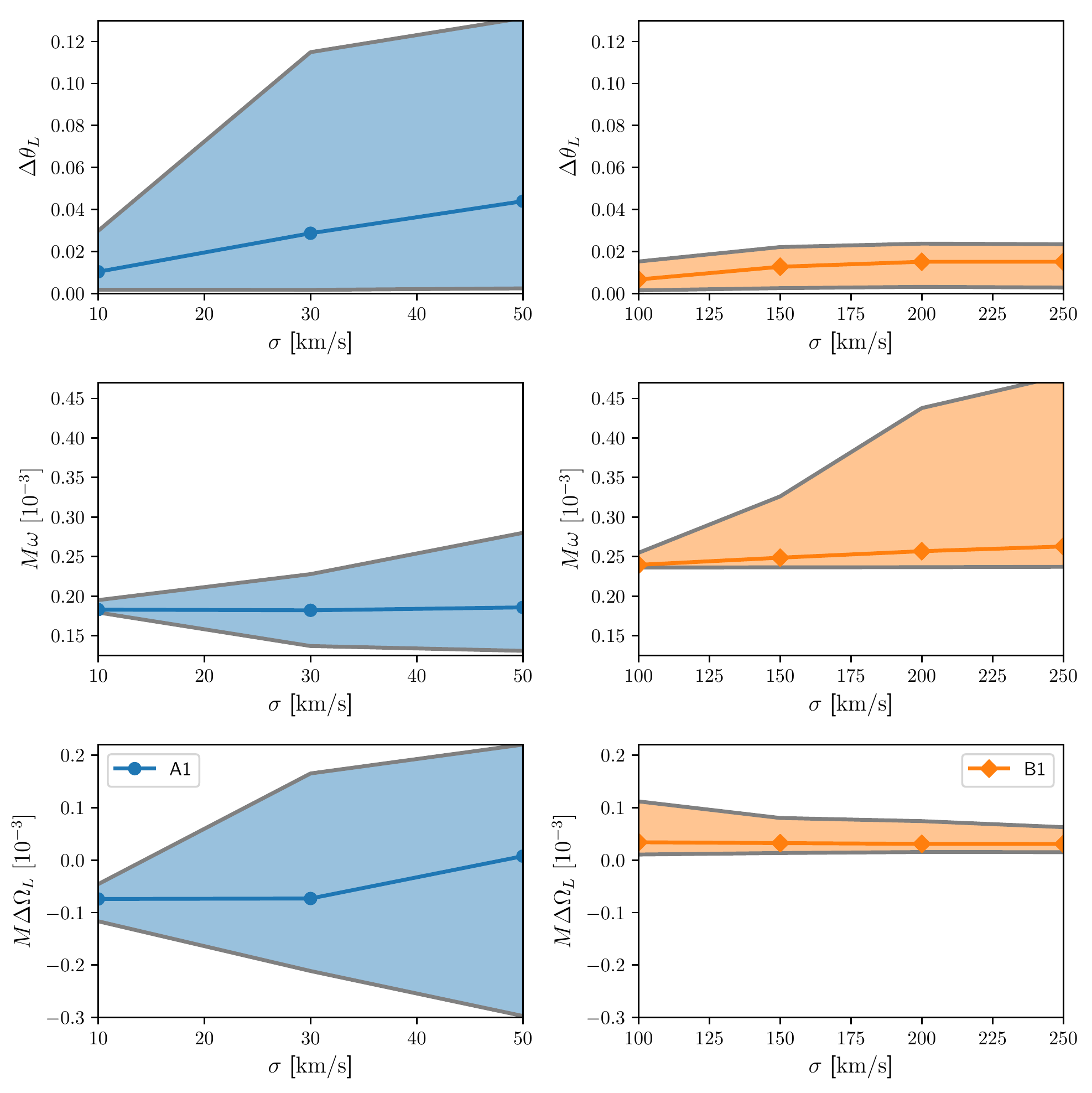}
   \caption{
   The dependence of the nutation amplitude $\Delta\theta_L$, nutation frequency $\omega$, and precession-frequency variation $\Delta\Omega_L$ on the natal kick velocity dispersion $\sigma$, respectively, for BBHs that originate from Pathway A1 (B1), i.e., SMT1-SN1-CEE2-SN2 (CEE1-SN1-SMT2-SN2), assuming weak core-envelope coupling and negligible mass loss in BH formation due to the Kerr limit. The stellar binaries are initialized with the same parameter values as in Fig.~\ref{F:VsSigma}.
   The blue (orange) region indicates 90\% of BBHs in each distribution that evolved from Pathway A1 (B1), and the solid colored line marked by circles (diamonds) is the median percentile.  
   } \label{F:NutVsSigma}
\end{figure*}

The strength of natal kicks, parameterized by $\sigma$, that is needed to produce appreciable spin-orbit misalignments depends on the pathways of evolution shown in Fig.~\ref{F:Diagram}. If alignment due to tides or accretion is avoided, then BBH spin precession is possible, and if the spin magnitude of the primary BH is high, i.e. $\chi \gtrsim 0.5$, then the precession of $\mathbf{L}$ can be significant. Additionally, if the secondary BH spin is high, then spin-spin coupling is important near the end of the BBH inspiral, allowing for appreciable nutation of $\mathbf{L}$. These are the necessary ingredients to obtain precessing and nutating BBHs, which depend on the initial stellar binary parameters and assumptions, as we demonstrate below.

\subsection{Spin Precession Parameters}
\label{subsec:ResultsParams}

The dependence of the spin precession parameters of our isolated-channel BBHs on the initial stellar binary parameters are presented in this subsection. As stated in the figure captions, different initial assumptions are chosen to highlight signatures of precession or nutation. We use a consistent coloring and marking scheme that correspond to the pathway of evolution: blue circles, red pluses, orange diamonds, and green crosses correspond to Pathways A1, A2, B1, and B2, respectively.

Figure~\ref{F:VsSigma} depicts the dependence of the aligned effective spin $\chi_{\rm eff}$, the precession amplitude $\langle\theta_L\rangle$, and the precession frequency $\langle\Omega_L\rangle$ on the natal-kick strength parameter $\sigma$. BBHs in the left hand side plots evolve in Pathway A1, which requires smaller values of $\sigma$ to avoid unbinding all of the binaries in the SN kick of the primary, where we choose an $a_{\rm ZAMS}$ sufficiently large to suppress tidal alignment. BBHs in the right hand side evolve in Pathway B1 with larger values of $\sigma$. In both cases, we assume weak core-envelope coupling, but we assume a larger value of $f_{\rm B}$ in A1 than in B1 to demonstrate the similarities of the precession parameters between an example of generic precession (A1) and an example of regular precession (B1), since both BHs in A1 have high spins but only the primary in Pathway B1 has a high spin from accretion. This difference explains why the 5th percentile line (lower grey curve) for $\chi_{\rm eff}$ reaches lower values in the top left panel (A1) of Figure~\ref{F:VsSigma} than the top right panel (B1). In both Pathways A1 and  B1, the median of $\langle \theta_L \rangle$ increases and the medians of $\chi_{\rm eff}$ and $\langle\Omega_L\rangle$ decrease monotonically as $\sigma$ increases, because larger misalignments correspond to smaller values of $\cos\theta_i$ and $J$. The dependence of the precession parameters on $\sigma$ for Pathways A2 and B2 is similar to the dependence shown for Pathway B1. 

Figure~\ref{F:NutVsSigma} depicts the dependence of the nutation amplitude $\Delta\theta_L$, nutation frequency $\omega$, and precession-frequency variation $\Delta\Omega_L$ on the natal-kick dispersion $\sigma$ for the same parameter choices depicted in Figure~\ref{F:VsSigma}. The nutation amplitude increases with $\sigma$ in both pathways due to the greater spin misalignments, although it is smaller by a factor $\sim 4$ in Pathway B1 because the secondary inherits a small natal spin ($f_{\rm B} = 0.01$) and nutation is a consequence of spin-spin coupling. Since the orbital velocity prior to the second natal kick remains roughly constant, the median of $\Delta\theta_L$ in B1 plateaus in the limit of large $\sigma$ as the medians of the cosines of the primary and secondary misalignments plateau at $\sim 0.8$ for these chosen values of $m_{1,\rm ZAMS},~m_{2,\rm ZAMS},$ and $a_{\rm ZAMS}$, e.g., see the right panel of Fig. 6 of \cite{Steinle2021}).

The median of the nutation frequency $\omega$ in A1 is approximately independent of $\sigma$ because the mass ratio $q \approx 0.83$ and the spin magnitudes ($\chi_1 \approx 0.79 < \chi_2 \approx 0.86$) are independent of $\sigma$.  As only a small fraction of binaries experience tidal alignment of the secondary, the median values of $\cos\theta_{1\infty}$ and $\cos\theta_{2\infty}$, which are the misalignments as $r/M \to \infty$, decrease with $\sigma$ while their scatter increases. This increases the scatter in $\omega$ while leaving its median largely unchanged as seen in the third column and third row of Fig. 4 of \cite{GangardtSteinle2021}.

The mass ratio $q \approx 0.54$ is much smaller in Pathway B1 than A1, because SMT from the secondary to the primary reduces the mass ratio below $q_{\rm ZAMS} = 5/7$ rather than enhancing it.  This implies larger median values of $\omega$ in B1 since $\omega \propto (1-q)/(1+q)$ at lowest PN order.  The larger primary spin in B1 ($\chi_1 \approx 0.69 > \chi_2 \approx 0.17$) due to this same accretion onto the primary causes $\omega$ to increase with $\sigma$ as $\cos\theta_{1\infty}$ decreases as seen in the second column and third row of Fig. 4 of \cite{GangardtSteinle2021}.

The behavior of the precession-frequency variation $\Delta\Omega_L$ can be understood by considering the $\cos\theta_1-\cos\theta_2$ plane at $r/M \to \infty$. The median of $\Delta\Omega_L$ in A1 is less than zero for small $\sigma$, because $q \approx 0.83$, the secondary BH spin magnitude is larger than the primary BH spin magnitude, and most binaries are located in the upper-right corner of the $\cos\theta_{1\infty}-\cos\theta_{2\infty}$ plane with a preference for $\cos\theta_1 < \cos\theta_2$ as the secondary WR star can be tidally aligned prior to SN2, analogous to the panel in the fifth column and fourth row of Fig. 4 of \cite{GangardtSteinle2021}. As $\sigma$ increases to large values, the median of $\Delta\Omega_L \sim 0$ as the natal kick of the secondary creates more scatter along the 1:1 diagonal in the $\cos\theta_{1\infty}-\cos\theta_{2\infty}$ plane with a preference for $\cos\theta_1 < \cos\theta_2$. In B1, $\Delta\Omega_L$ is strictly positive and approximately constant with $\sigma$, because $q \approx 0.54$ and only the primary has a high spin magnitude, analogous to the second column and fourth row of Fig. 4 of \cite{GangardtSteinle2021}. As $\sigma$ increases, the scatter in $\Delta\Omega_L$ decreases as fewer binaries are near the boundary defined by $\mathbf{J}\parallel\mathbf{L}$ in the $\cos\theta_{1\infty}-\cos\theta_{2\infty}$ plane where $\Delta\Omega_L \to \pm\infty$.

The main effect of $\sigma$ is to change the variance in the distributions of the spin precession parameters. This is consistent with Fig. 4 of Gangardt \& Steinle {\it et al.} 2021 where these parameters change most significantly in the $\cos\theta_{1\infty}-\cos\theta_{2\infty}$ plane in the vicinity of the $\mathbf{J}\parallel\mathbf{L}$ boundary. As binaries with two high spins are most likely to be in the proximity of the $\mathbf{J}\parallel\mathbf{L}$ boundary, the precession parameters for binaries in Pathway A1 with two high spins have a stronger dependence on $\sigma$ than those in B1 with only one high spin. The precession $\langle\Omega_L\rangle$ and nutation $\omega$ frequencies depend weakly on the asymptotic misalignments as they are spin-independent at leading PN order (see Sec.~II.C. of Gangardt \& Steinle {\it et al.} 2021).

The fraction $f_{\rm merge}$ of binaries that merge within the age of the Universe after the second natal kick is roughly constant in $\sigma$, i.e., $f_{\rm merge} \sim$ 0.7 (0.4) in A1 (B1), but decreases in the large-$\sigma$ limit. The larger $a_{\rm ZAMS}$ and $\sigma$ yields larger $f_{\rm merge}$ in B1 than in A1, but the later onset of CEE in A1 causes fewer binaries to survive the first natal kick than in B1. A complicated interplay between the occurrence of CEE relative to the natal kicks and larger post-kick semimajor axes and eccentricities (which are competing effects in $t_{\rm merge}$) from stronger kicks (i.e., larger $\sigma$) cause a slight dependence of $f_{\rm merge}$ on $\sigma$.

\begin{figure*}[!t] 
  \centering
  \includegraphics[width=1.0\linewidth]{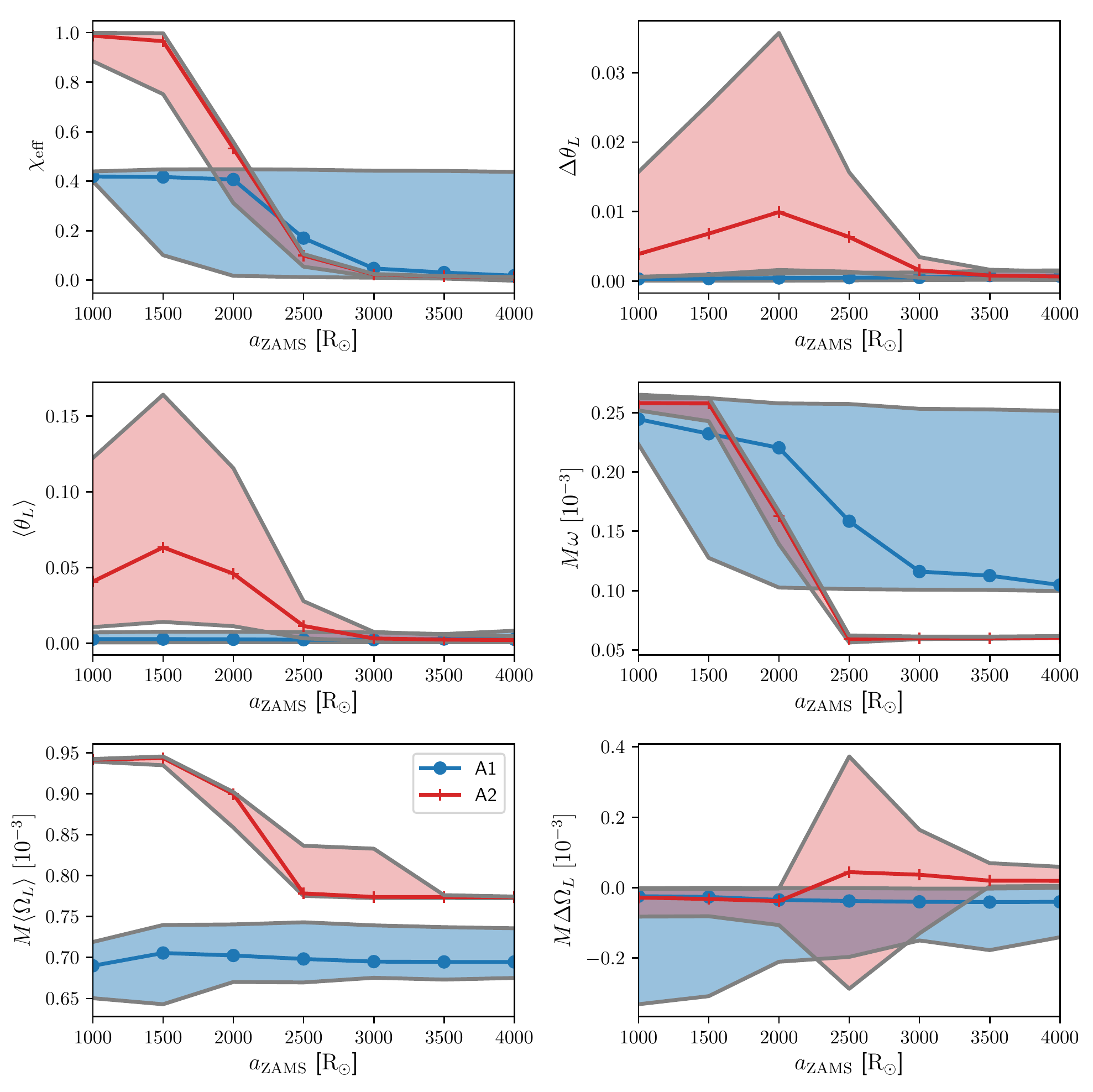}
   \caption{The dependence of the aligned effective spin $\chi_{\rm eff}$, average precession amplitude $\langle\theta_L\rangle$, and average precession frequency $\langle\Omega_L\rangle$, the nutation amplitude $\Delta\theta_L$, the nutation frequency $\omega$, and the precession variation $\Delta\Omega_L$ on the zero-age main sequence binary separation $a_{\rm ZAMS}$, respectively, for BBHs that originate from Pathway A1 (A2), i.e., SMT1-SN1-CEE2-SN2 (SMT1-CEE2-SN1-SN2), assuming strong core-envelope coupling and isotropic mass-loss in BH formation due to the Kerr limit. The stellar binaries are initialized with $Z = 0.0002$, $m_{1,\rm ZAMS} = 70$~M$_{\odot}$, $f_{\rm a} = 0.2$, $m_{2,\rm ZAMS} = 50$~M$_{\odot}$ ($m_{2,\rm ZAMS} = 67$~M$_{\odot}$), and $\sigma = 30$~km/s ($\sigma = 200$~km/s). The blue (red) region indicates 90\% of BBHs in each distribution that evolved from Pathway A1 (A2), and the solid colored line marked by circles (pluses) is the median percentile. 
   } \label{F:VsAzams}
\end{figure*}

Figure~\ref{F:VsAzams} shows the dependence of the aligned effective spin $\chi_{\rm eff}$, the precession amplitude $\langle\theta_L\rangle$, the precession frequency $\langle\Omega_L\rangle$, the nutation amplitude $\Delta\theta_L$, the nutation frequency $\omega$, and the precession-frequency variation $\Delta\Omega_L$ on the initial binary separation $a_{\rm ZAMS}$ for BBHs that evolve in Pathway A1 (blue circles) or in Pathway A2 (red pluses). Unlike in Figs.~\ref{F:VsSigma} and \ref{F:NutVsSigma}, we now assume strong core-envelope coupling to assess the effect of tides on the spin-precession parameters. This leads to small initial WR spins ($\chi \sim 0.001$) and thus small precession and nutation amplitudes at wide separations ($a_{\rm post-CE} \gtrsim 20$~R$_{\odot}$) at which tides are typically ineffective in both pathways.

In Pathway A1, the spin of the secondary WR star typically experiences tidal synchronization and alignment for $a_{\rm ZAMS} \lesssim 3{,}000$~R$_{\odot}$, increasing $\chi_{\rm eff}$ and $\omega$ at these separations.  Even at larger separations, the randomness of the first natal kick creates a $\gtrsim 5\%$ subpopulation with small enough semi-major axes for tidal synchronization and alignment of the secondary WR star to occur.  The small kick magnitude ($\sigma$ = 30 km/s) implies that only negligible misalignments can be generated by the second natal kick, leading to unobservable precession in this pathway ($\langle\theta_L\rangle \lesssim 0.01$).  Nutation is even further suppressed by the negligible primary spin, since the primary collapses prior to CEE which shrinks the binary separation to values at which tidal synchronization can occur.

In Pathway A2, the BBH mass ratio $q \approx 0.9$ is constant with respect to $a_{\rm ZAMS}$, as there is no scatter in the masses due to both mass-transfer events occurring before either SN events. Both WR stars experience tides for small $a_{\rm ZAMS}$ during the double-WR phase after CEE of the secondary, and the secondary WR star experiences tides again between the two natal kicks.
This allows both spins to be tidally synchronized and aligned, producing larger $\chi_{\rm eff}$ and $\langle\Omega_L\rangle$ than in A1. As $a_{\rm ZAMS}$ increases, the tidal torque diminishes and $\chi_{\rm eff}$ and $\langle\Omega_L\rangle$ asymptote to the values corresponding to the small spin magnitudes resulting from strong core-envelope coupling.
The precession and nutation amplitudes, $\langle\theta_L\rangle$ and $\Delta\theta_L$, respectively, are greater in A2 than A1 for $a_{\rm ZAMS} \lesssim 2{,}500$~R$_{\odot}$ because the high spin magnitudes of both WR stars, created by tidal synchronization following CEE, can become misaligned by the large second natal kick ($\sigma$ = 200 km/s).

The unusual dependence of the precession-frequency variation $\Delta\Omega_L$ on the initial separation $a_{\rm ZAMS}$ in Pathway A2 can be understood by considering the $\cos\theta_1-\cos\theta_2$ plane at $r/M \to \infty$. The boundary separating positive and negative values of $\Delta\Omega_L$ in the $\cos\theta_{1\infty}-\cos\theta_{2\infty}$ plane is determined by the condition $\mathbf{J}\parallel\mathbf{L}$ and depends on the spin magnitudes. Tidal synchronization leads to high ($\chi \gtrsim 0.5$), moderate ($\chi \sim 0.1$), and low ($\chi \lesssim 0.01$) spins for $a_{\rm ZAMS} \lesssim$ 1,500 R$_{\odot}$, $\sim 2,500$ R$_{\odot}$, and $\gtrsim$ 3,500 R$_{\odot}$ respectively.
We find that the $\mathbf{J}\parallel\mathbf{L}$ boundary for high spins ($a_{\rm ZAMS} \lesssim$ 1,500 R$_{\odot}$) closely resembles that shown in the panel in the fifth column and fourth row of Fig.~4 of \cite{GangardtSteinle2021}, implying that highly aligned binaries in the upper right corner will have $\Delta\Omega_L < 0$.  Although Fig.~4 of \cite{GangardtSteinle2021} only explored high spins, moderate and low spins yield $\mathbf{J}\parallel\mathbf{L}$ boundaries analogous to those with smaller mass ratios in the fourth and second columns of Fig.~4 of \cite{GangardtSteinle2021}.  These boundaries migrate to the right edge of the $\cos\theta_{1\infty}-\cos\theta_{2\infty}$ plane with smaller spins (larger $a_{\rm ZAMS}$).  For moderate spins ($\chi \sim 0.1$, $a_{\rm ZAMS} \sim$ 2,500 R$_{\odot}$), the binaries straddle the $\mathbf{J}\parallel\mathbf{L}$ boundary (where $\Delta\Omega_L \to \pm\infty$) leading to the large observed scatter in $\Delta\Omega_L$ at this separation.  For low spins ($\chi \lesssim 0.01$, $a_{\rm ZAMS} \gtrsim$ 3,500 R$_{\odot}$), the binaries all lie to the left of the $\mathbf{J}\parallel\mathbf{L}$ boundary implying $\Delta\Omega_L > 0$ at these separations.

The precession and nutation amplitudes $\langle\theta_L\rangle$ and $\Delta\theta_L$ both have maxima as functions of $a_{\rm ZAMS}$ in A2, because of the competition between tidal synchronization and alignment. This contrasts with $\chi_{\rm eff}$ for which tidal synchronization and alignment cooperate to produce high values at small $a_{\rm ZAMS}$.
In both Pathways A1 and A2, the dependencies of the nutation frequency $\omega$ and the aligned effective spin $\chi_{\rm eff}$ on $a_{\rm ZAMS}$ are strongly correlated because of tidal synchronization; smaller $a_{\rm ZAMS}$ provide larger spin magnitudes and generally larger values of both $\chi_{\rm eff}$ and $\omega$.  This is consistent with the bottom-left panel of Fig. 3 of \cite{GangardtSteinle2021} in the regime of moderate-to-high mass ratio (in A1 $q \approx 0.75$, in A2 $q \approx 0.9$). The scatter in $\chi_{\rm eff}$ and $\omega$ is larger in A1 than in A2 due to the subpopulation of binaries that originate from small post-CEE separations, which does not occur in A2.

For the binaries in Fig.~\ref{F:VsAzams}, we assumed an accreted fraction of $f_{\rm a} = 0.2$, but the correlation between $\chi_{\rm eff}$ and $\omega$ is insensitive to this choice. If we instead assume $f_{\rm a} = 0$, there would not be MRR in A2, and the smaller (larger) BBH mass ratio $q$ in A1 (A2) would lead to higher (lower) values of $\omega$, as $\omega \propto (1-q)/(1+q)$ at lowest PN order. Higher $f_{\rm a}$ would increase (decrease) the BBH mass ratio in Pathways A1 (A2) due to MRR in Pathway A2, leading to a corresponding decrease (increase) in $\omega$.

Our results suggest that precession and nutation are undetectable in Pathway A1 due to the small amplitudes $\langle\theta_L\rangle$ and $\Delta\theta_L$ that result from strong core-envelope coupling. Natal kicks in the supernova of the secondary with $\sigma = 30$ km/s are too weak to misalign the secondary at separations small enough for it to acquire nonzero spin magnitude from tidal synchronization, implying that a larger value of $\sigma$ in the second natal kick might produce a larger precession amplitude $\langle\theta_L\rangle$ in A1. Although larger $\Delta\theta_L \sim 0.01$ is accessible in A2 as tidal synchronization provides high spin magnitudes and the secondary natal kick provides misalignments after tidal alignment, nutation of this amplitude would still be very difficult to detect in gravitational-wave observations. The inability of tides to produce BBHs with significant nutation supports our contention that high natal BH spin magnitudes are critical for observable nutation in the isolated channel.

\begin{figure}[!t] 
  \centering
  \includegraphics[width=1.0\linewidth]{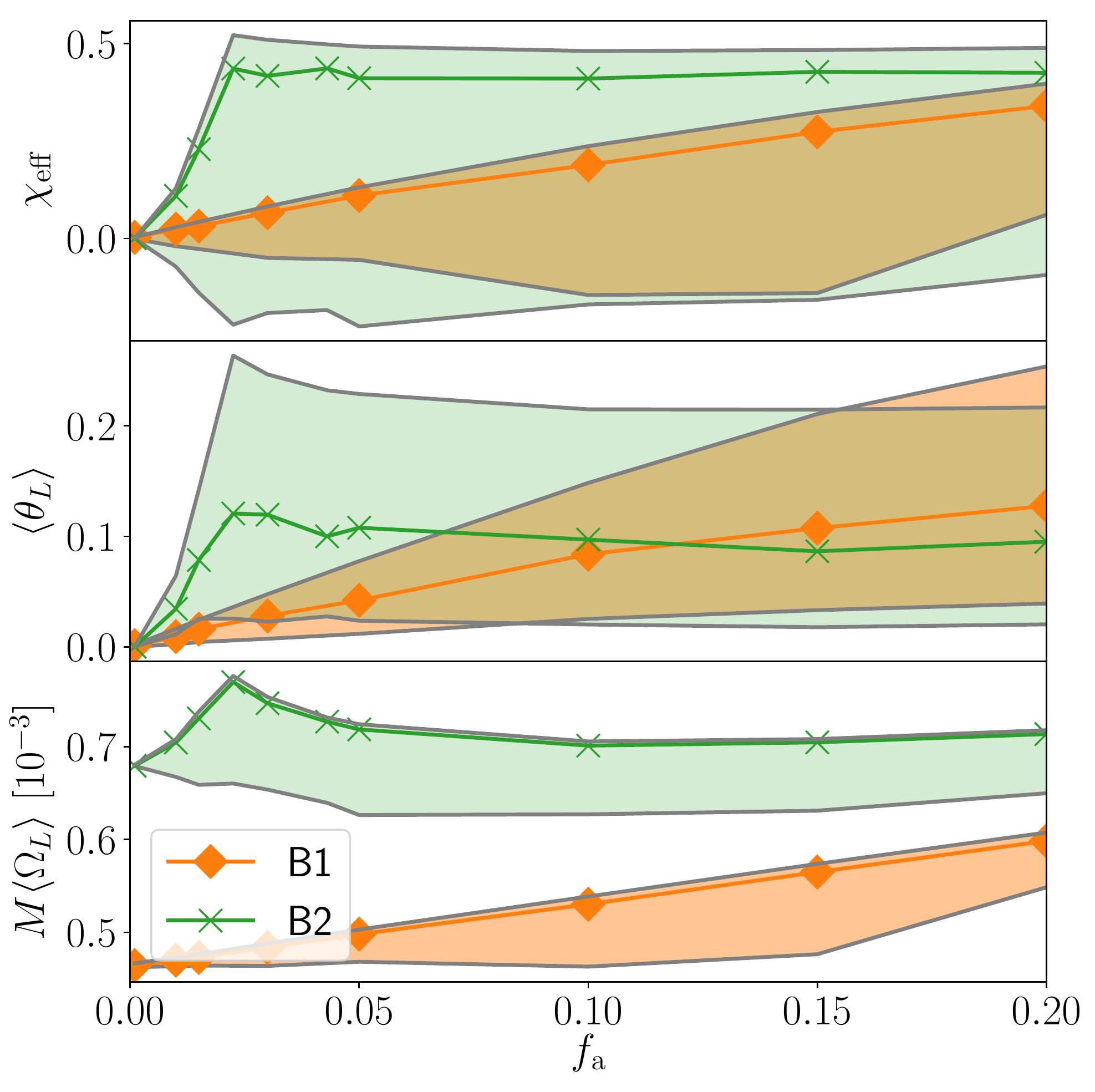}
   \caption{The dependence of the aligned effective spin $\chi_{\rm eff}$, average precession amplitude $\langle\theta_L\rangle$, and average precession frequency $\langle\Omega_L\rangle$ on the fraction of mass that is accreted during SMT $f_{\rm a}$, respectively, for BBHs that originate from Pathway B1 (B2), i.e., CEE1-SN1-SMT2-SN2 (CEE1-SMT2-SN1-SN2), assuming strong core-envelope coupling and isotropic mass loss in BH formation due to the Kerr limit. The stellar binaries are initialized with $Z = 0.0002$, $a_{\rm ZAMS} = 12{,}000$~R$_{\odot}$, $m_{1,\rm ZAMS} = 70$~M$_{\odot}$, $m_{2,\rm ZAMS} = 50$~M$_{\odot}$ ($m_{2,\rm ZAMS} = 67$~M$_{\odot}$), and $\sigma = 200$~km/s. The orange (green) region indicates 90\% of BBHs in each distribution that evolved from Pathway B1 (B2), and the solid colored line marked by diamonds (crosses) is the median percentile.
   }\label{F:VsAcc}
\end{figure}

Figure~\ref{F:VsAcc} depicts the dependence of the aligned effective-spin $\chi_{\rm eff}$, the precession amplitude $\langle\theta_L\rangle$, and the precession frequency $\langle\Omega_L\rangle$ on the fraction $f_{\rm a}$ of the secondary's envelope that is accreted by the primary in stable mass transfer as a BH in Pathway B1 (orange diamonds) and as a WR star in Pathway B2 (green crosses). In B1, the BBH mass ratio $q$ monotonically decreases with increasing $f_{\rm a}$ as the primary BH accretes an amount of mass proportional to the mass of the secondary's envelope and does not experience further subsequent mass loss. The aligned effective spin $\chi_{\rm eff}$, the precession amplitude $\langle\theta_L\rangle$, and the precession frequency $\langle\Omega_L\rangle$ each increase monotonically with $f_{\rm a}$ since the dimensionless spin of the primary BH increases monotonically with $f_{\rm a}$. 

\begin{figure*}[!t] 
  \centering
  \includegraphics[width=1.0\linewidth]{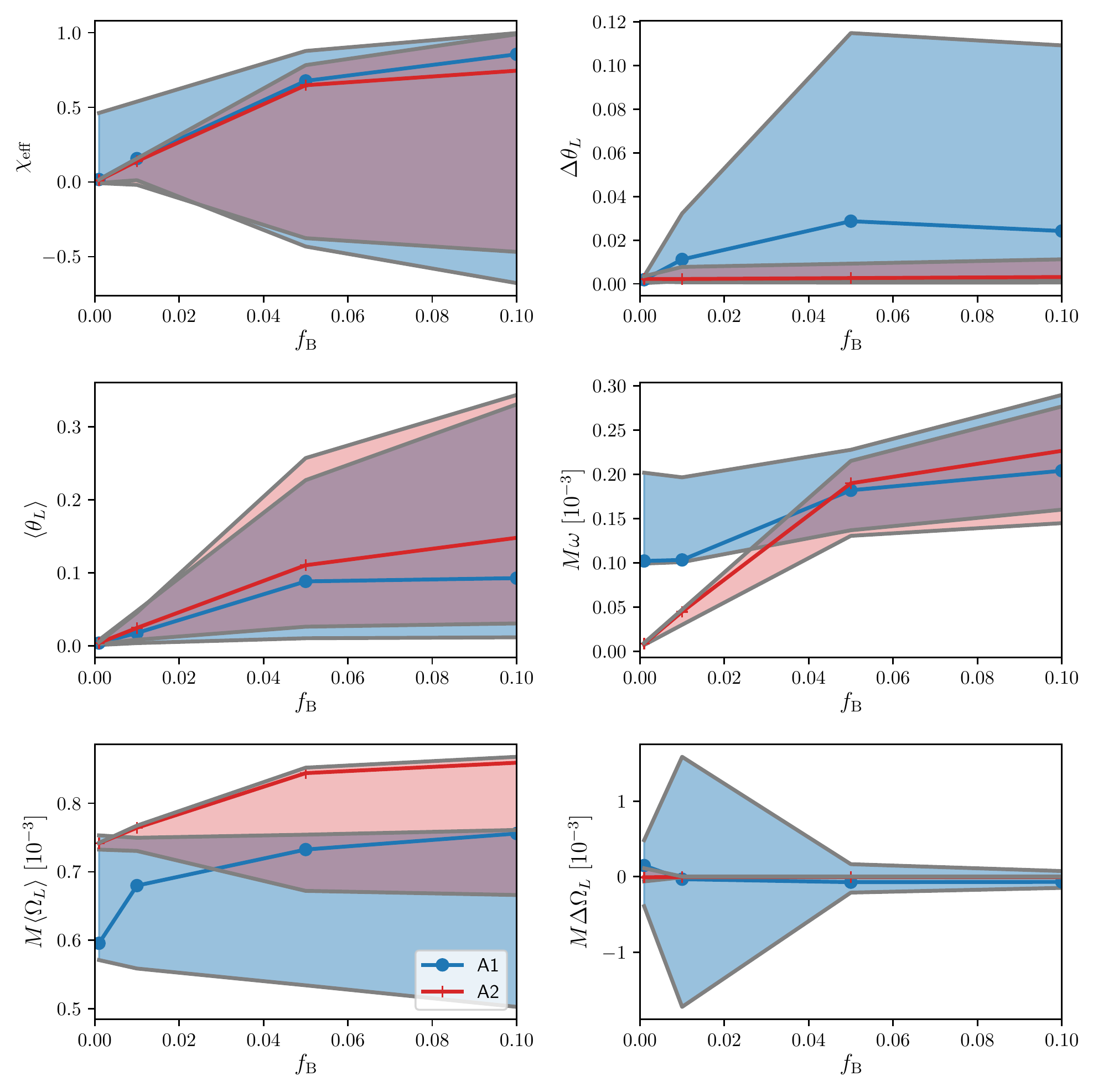}
   \caption{The dependence of the aligned effective spin $\chi_{\rm eff}$, average precession amplitude $\langle\theta_L\rangle$, and average precession frequency $\langle\Omega_L\rangle$, the nutation amplitude $\Delta\theta_L$, the nutation frequency $\omega$, and the precession variation $\Delta\Omega_L$ on the breakup fraction of the initial Wolf-Rayet stellar spins $f_{\rm B}$, respectively, for BBHs that originate from Pathway A1 (A2), i.e., SMT1-SN1-CEE2-SN2 (SMT1-CEE2-SN1-SN2), assuming weak core-envelope coupling and negligible mass-loss in BH formation due to the Kerr limit. The stellar binaries are initialized with $Z = 0.0002$, $a_{\rm ZAMS} = 6{,}000$~R$_{\odot}$, $m_{1,\rm ZAMS} = 70$~M$_{\odot}$, $f_{\rm a} = 0.2$, $m_{2,\rm ZAMS} = 50$~M$_{\odot}$ ($m_{2,\rm ZAMS} = 60$~M$_{\odot}$), and $\sigma = 30$~km/s ($\sigma = 200$~km/s). The blue (red) region indicates 90\% of BBHs in each distribution that evolved from Pathway A1 (A2), and the solid colored line marked by circles (pluses) is the median percentile.} \label{F:VsBreakA}
\end{figure*}

\begin{figure*}[!t] 
  \centering
  \includegraphics[width=1.0\linewidth]{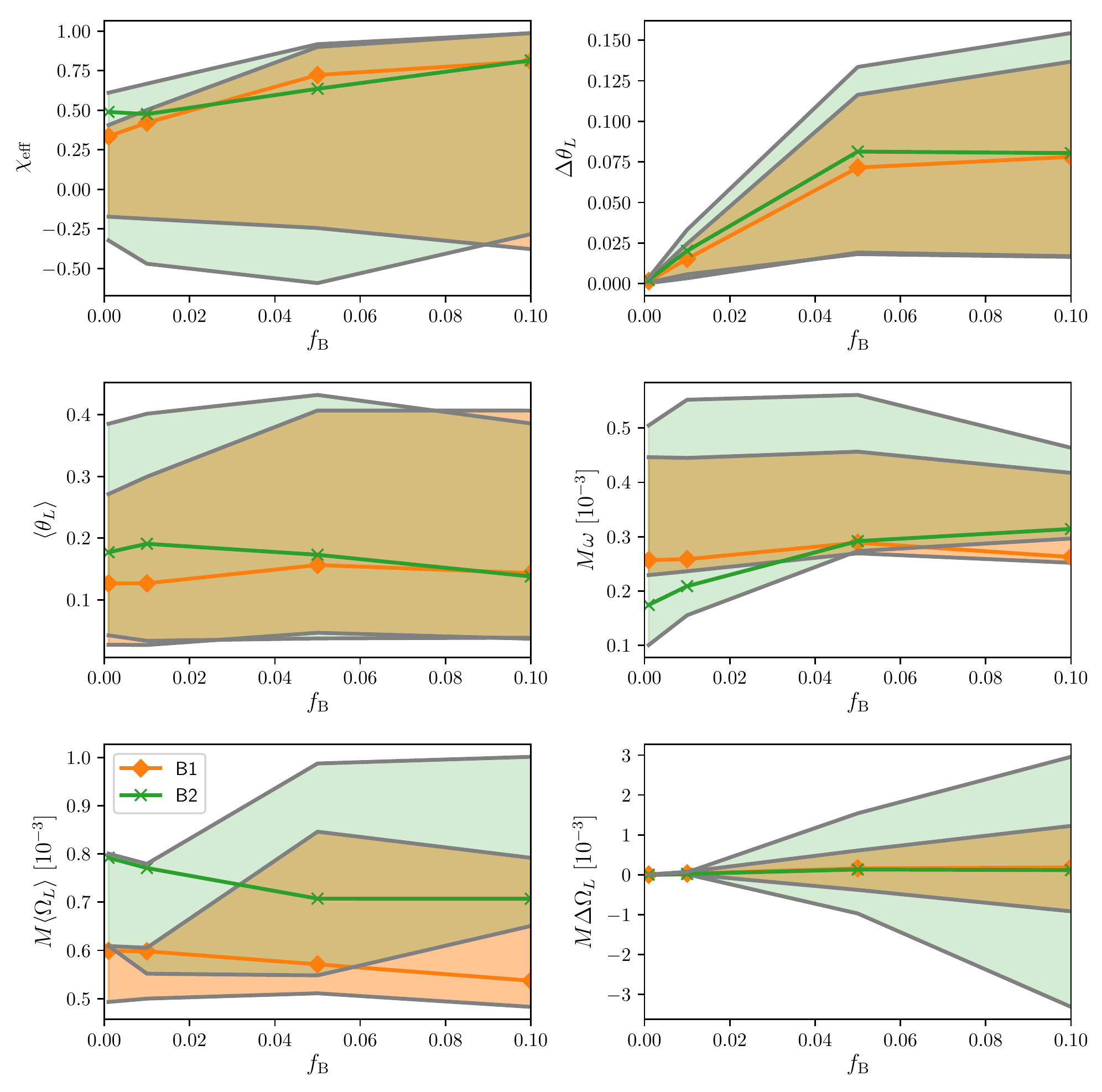}
   \caption{The dependence of the aligned effective spin $\chi_{\rm eff}$, average precession amplitude $\langle\theta_L\rangle$, average precession frequency $\langle\Omega_L\rangle$, nutation amplitude $\Delta\theta_L$, nutation frequency $\omega$, and precession variation $\Delta\Omega_L$ on the breakup fraction of the initial Wolf-Rayet stellar spins $f_{\rm B}$, respectively, for BBHs that originate from Pathway B1 (B2), i.e., CEE1-SN1-SMT2-SN2 (CEE1-SMT2-SN1-SN2), assuming weak core-envelope coupling and negligible mass loss in BH formation due to the Kerr limit. The stellar binaries are initialized with $Z = 0.0002$, $a_{\rm ZAMS} = 15{,}000$~R$_{\odot}$, $m_{1,\rm ZAMS} = 70$~M$_{\odot}$, $f_{\rm a} = 0.2$, $m_{2,\rm ZAMS} = 50$~M$_{\odot}$ ($m_{2,\rm ZAMS} = 60$~M$_{\odot}$), and $\sigma = 200$~km/s. The orange (green) region indicates 90\% of BBHs in each distribution that evolved from Pathway B1 (B2), and the solid colored line marked by diamonds (crosses) is the median percentile.} \label{F:VsBreakB}
\end{figure*}

The very efficient accretion by the primary WR star in Pathway B2 introduces the possibility of MRR if the WR star is spun up above the Kerr limit. We can estimate the minimum value of $f_{\rm a}$ at which this occurs by equating the accreted angular momentum $S_{\rm acc} \sim f_{\rm a} m_{\rm env}(mR)^{1/2}$ to that of a maximally spinning BH ($S_{\rm max} = m^2$) yielding $f_{\rm a,min} \sim (m/m_{\rm env})(m/R)^{1/2}$.
For $m/m_{\rm env}$ of order unity and $m \sim 50~M_\odot$, $R \sim R_\odot$, we have $f_{\rm a,min} \sim 10^{-2}$.
This estimate agrees with the location of the peak at
$f_{\rm a,min} \approx 0.023$ in the median values (green solid lines) of $\chi_{\rm eff}$, $\langle\theta_L\rangle$, and $\langle\Omega_L\rangle$ in Figure~\ref{F:VsAcc}. Below this value, $q$ decreases and the primary spin increases similar to Pathway B1. As $f_{\rm a} \to f_{\rm a,min}$, the final dimensionless spin of the primary WR star is $\chi \approx 0.9$, which is driven to unity by neutrino emission during core collapse. MRR does not occur until $f_{\rm a} = f_{\rm a,MRR} \approx 0.043$. For $f_{\rm a} > f_{\rm a,min}$, angular momentum must be lost during core collapse to produce a BH with spin below the Kerr limit.
As this angular momentum is carried by isotropic winds with lower specific angular momentum than the equatorial accretion flow, the accreted mass above $f_{\rm a,min}m_{\rm env}$ leads to a reduction in the mass of the BH formed from the binary. For $f_{\rm a} = f_{\rm a,MRR}$, this reduction leads to a primary BH mass equal to the secondary mass ($q = 1$). Above this value, MRR occurs and $q$, defined as the ratio of the lighter to heavier BH masses, again decreases with $f_{\rm a}$. As seen in the middle row of Fig.~3 of \cite{GangardtSteinle2021}, $\langle\theta_L\rangle$ and $\langle\Omega_L\rangle$ are insensitive to the mass ratio near $q = 1$ consistent with the weak dependence of these parameters for $f_{\rm a} 
\approx f_{\rm a,MRR}$.

Figures~\ref{F:VsBreakA} and \ref{F:VsBreakB} show the dependence of the aligned effective-spin $\chi_{\rm eff}$, the precession amplitude $\langle\theta_L\rangle$, the precession frequency $\langle\Omega_L\rangle$, the nutation amplitude $\Delta\theta_L$, the nutation frequency $\omega$, and the precession-frequency variation $\Delta\Omega_L$ on the Wolf-Rayet breakup-spin fraction $f_{\rm B}$ in Scenarios A and B, respectively.  For simplicity, we assume the same value of $f_{\rm B}$ for both natal WR stars in each binary. As in Fig.~\ref{F:VsAzams}, the BBHs of Pathway A1 exhibit scatter in $\chi_{\rm eff}$, $\langle\Omega_L\rangle$, $\omega$, and $\Delta\theta_L$ even in the limit $f_{\rm B} \to 0$ because of the subpopulation with small semi-major axes following the first natal kick that experience tidal effects. The nutation amplitude $\Delta\theta_L$ is greatly suppressed in A2 compared to in A1 because accretion onto the secondary star during SMT results in a near unity BBH mass ratio ($q \approx 0.98$).

For the binaries from Scenario A in Fig.~\ref{F:VsBreakA}, the precession and nutation parameters all monotonically increase with $f_{\rm B}$ because the spin magnitudes of the BHs are purely determined by the value of $f_{\rm B}$. $\langle\Omega_L\rangle$ increases with $f_{\rm B}$ as both BH spin magnitudes increase, but the misalignments and the mass ratio are constant with $f_{\rm B}$. Even though we allow for modest accretion ($f_{\rm a} = 0.2$), this does not directly effect the spin magnitude of the secondary BH since the spin that the secondary star gains on the main sequence in SMT is not inherited by its core as a Wolf-Rayet star. Accretion indirectly effects the spin of the secondary BH as its progenitor's gain in mass results in a more massive secondary WR star and hence a smaller dimensionless natal WR spin for any value of $f_{\rm B}$ (see Fig. 3 of \cite{Steinle2021}), but this effect is insignificant unless we assume a much larger amount of accretion ($f_{\rm a} \gg 0.5$). For $f_{\rm B} \lesssim 0.01$, the nutation frequency $\omega$ is generally smaller in A2 than in A1 since the mass ratio is larger in A2, $\omega \propto (1-q)/(1+q)$, and the dimensionless spin magnitudes are very small ($\chi \approx 0.01$). As $f_{\rm B}$ increases, the nearly equal BH spins dominate over the dependence on the mass ratio and $\omega$ is comparable in both pathways.

The precession-frequency variation $\Delta\Omega_L$, shown for Senario A in the bottom right panel of Fig.~\ref{F:VsBreakA}, has enormous scatter in a narrow band near $f_{\rm B} \approx 0.01$ for Pathway A1.  This is similar to the large scatter for the same parameter for Pathway A2 at $a_{\rm ZAMS} \approx 2{,}500$~R$_{\odot}$ in the bottom right panel of Fig.~\ref{F:VsAzams}.  The underlying cause is the same in both cases; for these parameter choices, the $\mathbf{J}\parallel\mathbf{L}$ boundary at which $\Delta\Omega_L \to \pm\infty$ passes through the location at which the BBHs are clustered in the $\cos\theta_{1\infty} - \cos\theta_{2\infty}$ plane. In Pathway A2, the near unity mass ratio $q \approx 0.98$ suppresses $\Delta\Omega_L$ for all values of $f_{\rm B}$.

The role of accretion is more significant in Scenario B as shown in Fig.~\ref{F:VsBreakB}, because it provides the primary a dimensionless spin $\chi \approx$ 0.6 ($\chi \approx 1$) in Pathway B1 (B2) even in the absence of natal spins ($f_{\rm B} = 0$). This leads to a larger precession amplitude $\langle\theta_L\rangle$ at small values of $f_{\rm B}$ for Scenario B than Scenario A. As $f_{\rm B}$ increases, the spin asymmetry between the primary and secondary is reduced and the $\mathbf{J}\parallel\mathbf{L}$ boundary migrates away from the right edge of the $\cos\theta_{1\infty} - \cos\theta_{2\infty}$ plane as can be seen by comparing the second and fourth columns of the fourth row in Fig. 4 of \cite{GangardtSteinle2021}.  More binaries near this boundary for $f_{\rm B} \gtrsim 0.01$ leads to greater scatter in $\Delta\Omega_L$, and more binaries on the far side of this boundary leads to an increase in the 95\% of $\langle\Omega_L\rangle$.  However, for the majority of binaries that remain near the $\mathbf{J}\parallel\mathbf{L}$ boundary, the reduction in the spin asymmetry with increasing $f_{\rm B}$ causes a decrease in the median value of $\langle\Omega_L\rangle$.  These effects are more pronounced in Pathway B2 than B1 because the primary acquires a maximal spin even for $f_{\rm B} = 0$ by accreting as a WR star during SMT, increasing the spin asymmetry between it and the secondary for all $f_{\rm B} < 0.1$.

Together, these results imply that inheritance of high spin via weak core-envelope coupling is crucial for BBHs to experience significant nutation. Although tidal synchronization can yield large spins for the secondary (and primary) in Pathway A1 (A2), tidal alignment will suppress nutation for all but the largest of secondary natal kicks. If these were associated with large primary kicks (as in the choice of parameters here), merger rates would be highly suppressed in Pathway A1 in which the primary natal kick occurs before CEE shrinks the binary separation.  Accretion onto the primary in Scenario B can yield significant precession provided that it is not Eddington limited and that the misalignment is not suppressed by the Bardeen-Petterson effect \cite{Bardeen1975} and an insufficiently large secondary natal kick. However, nutation will still be suppressed in Scenario B if the secondary cannot acquire its own spin, as it arises from spin-spin coupling.  Significant nutation can only occur if both BHs have moderate to large spins, and this can only occur if one or both BHs inherits its spin via weak core-envelope coupling, depending upon the evolutionary pathway.

\begin{figure*}[!t] 
  \centering
  \includegraphics[width=1.0\linewidth]{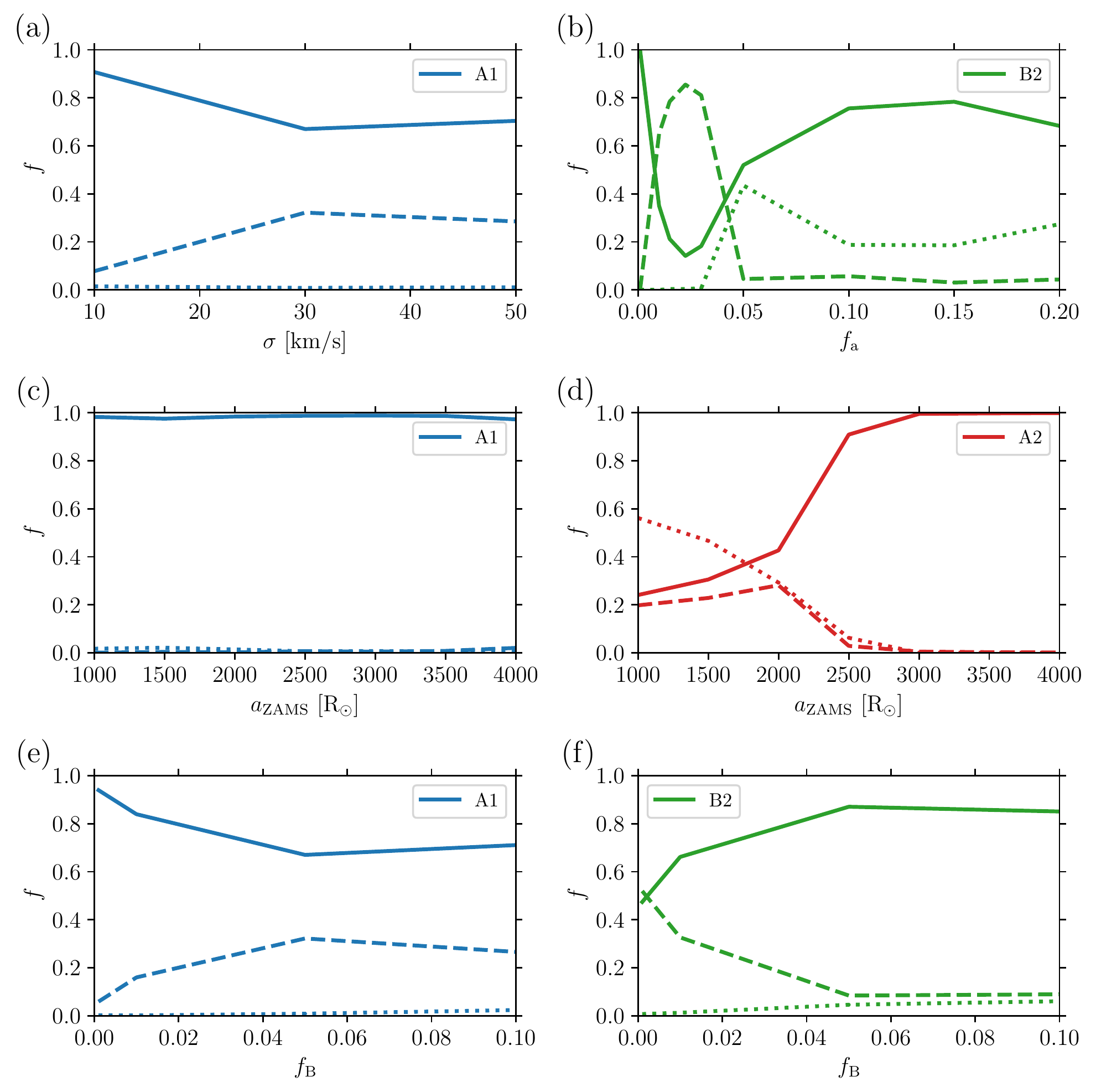}
   \caption{Fractions of spin-precession morphology, where the dashed, dotted, and solid lines correspond to the fractions of BBHs in the librating about $\pi$, librating about $0$, and circulating morphologies, respectively.
   Panels (a), (b), (c) and (d), (e), and (f) correspond to the BBHs depicted in Fig.'s~\ref{F:VsSigma}, \ref{F:VsAcc}, \ref{F:VsAzams}, \ref{F:VsBreakA}, and \ref{F:VsBreakB}, respectively. The binaries in panels (a), (c), and (e) assume the same ZAMS masses whereas the binaries in panel (d) assume a larger value of $m_{2, \rm ZAMS}$ to evolve in pathway 2, and the binaries in panel (b) (panel (f)) assume $m_{2, \rm ZAMS} = 67$ M$_{\odot}$ ($m_{2, \rm ZAMS} = 60$ M$_{\odot}$). Strong (weak) core-envelope coupling is assumed in panels (b), (c), and (d) (panels (a), (e), and (f)).
   } \label{F:Morphs}
\end{figure*}

\subsection{Spin Precession Morphologies}
\label{subsec:ResultsMorphs}

Another insight from the works of \citeauthor{Kesden2015}~\cite{Kesden2015} and \citeauthor{Gerosa2015b}~\cite{Gerosa2015b} was the identification of three spin precession $\emph{morphologies}$ that are characterized by the change over a nutation period of $\Delta\Phi$, the angle subtended by the projections of the two BH spins in the orbital plane (see Fig.~1 of \cite{Gerosa2015b}). 
The morphological classes are: $\Delta\Phi$ circulates through the full range $[-\pi,\pi]$, $\Delta\Phi$ librates about 0 but never reaches $\pm\pi$, and $\Delta\Phi$ librates about $\pm\pi$ but never reaches 0. For given BBH mass ratio, dimensionless spin magnitudes, and binary separation, the morphology is specified by $J$ and $\chi_{\rm eff}$, implying that the morphology evolves only on the radiation-reaction timescale and that they form disconnected regions in the $J-\chi_{\rm eff}$ (or equivalently, $\cos\theta_{1\infty} - \cos\theta_{2\infty}$) planes. These morphologies are useful for interpreting the astrophysical origin of BBHs detected by GWs \cite{Gerosa2013}.  The panels of Figure~\ref{F:Morphs} depict the three morphological fractions for the distributions of BBHs presented in Figures~\ref{F:VsSigma}\,-\,\ref{F:VsBreakB}.

Panel (a) of Fig.~\ref{F:Morphs} shows the morphology fractions as a function of the natal kick dispersion $\sigma$ in Pathway A1 with weak core-envelope coupling and natal WR spin $f_{\rm B} = 0.05$.  For this pathway, the spin misalignments are almost entirely determined by the primary natal kick which occurs before CEE reduces the binary separation.  For small $\sigma$, the BBHs are primarily in the circulating morphology consistent with equal asymptotic spin misalignments like the BBHs along the diagonal of the $\cos\theta_{1\infty} - \cos\theta_{2\infty}$ plane in the upper right panel of Fig.~13 of \cite{Gerosa2015b}.  As $\sigma$ increases, a higher fraction of binaries experience tidal realignment of the secondary, driving $\cos\theta_{2\infty} \to 1$ and moving binaries above the diagonal where the morphology librating about $\pi$ predominates.

In panel (b) of Fig.~\ref{F:Morphs}, where strong core-envelope coupling is assumed, the dependence of the morphology fractions on the accreted fraction $f_{\rm a}$ for BBHs from Pathway B2 shows complicated behavior due to the interplay of the BBH mass ratio, spins, and MRR. In the limit of zero accretion, all of the binaries are circulating since both black holes have negligible spin. As $f_{\rm a}$ increases to $f_{\rm a,min} \approx 0.023$ where the primary spin reaches the Kerr limit, coupling to this large primary spin traps $\sim80\%$ of binaries into the morphology librating about $\pi$ which dominates for these parameters (see the upper right corner of the bottom center panel of Fig.~14 of \cite{Gerosa2015b}). For $f_{\rm a} > f_{\rm a,MRR} \approx 0.043$, MRR occurs (the primary star forms the less massive BH).  When the less massive BH is more highly spinning, the upper right corner of the ($\cos\theta_{1\infty} - \cos\theta_{2\infty})$ plane is again dominated by the circulating morphology, although the morphology librating about $0$ also provides a significant contribution (see the bottom left panel of Fig.~14 of \cite{Gerosa2015b}).

Panel (c) of Fig.~\ref{F:Morphs} shows the morphology fractions for BBHs from Pathway A1 versus the initial binary separation $a_{\rm ZAMS}$ where we assume strong stellar core-envelope coupling and hence negligible natal spins. In contrast to panel (a), in which weak coupling and $f_{\rm B} = 0.05$ lead to two large misaligned spins and generic precession, here essentially all of the binaries are in the circulating morphology consistent with a single aligned spin. Panel (d) also shows the morphology fractions versus $a_{\rm ZAMS}$, but the kick dispersion is higher ($\sigma = 200$~km/s versus 30~km/s) and the higher initial secondary mass ($m_{2,\rm ZAMS} = 67$~M$_{\odot}$ versus $50$~M$_{\odot}$) implies Pathway A2 instead of A1.  At wide separations $a_{\rm ZAMS} \gtrsim$~3,000~R$_{\odot}$ where tides are ineffective, the binaries are all in the circulating morphology as in Pathway A1.  For smaller values of $a_{\rm ZAMS}$, tidal synchronization before core collapse of the primary star can spin up both stars. The primary spin is misaligned by both natal kicks, but the secondary can be tidally realigned after the first natal kick and on average receives a smaller misalignment from the second natal kick alone. However, the high value of $q_{\rm ZAMS}$ and the SMT from primary to secondary in Scenario A causes MRR to occur for all values of $a_{\rm ZAMS}$. This implies that the more massive BH is more aligned, the binaries cluster below the diagonal in the $\cos\theta_{1\infty} - \cos\theta_{2\infty}$ plane, and the morphology librating about $0$ is increasingly favored at small $a_{\rm ZAMS}$.

The bottom panels (e) and (f) of Fig.~\ref{F:Morphs} show the morphology fractions versus the WR breakup spin fraction $f_{\rm B}$ in Pathways A1 and B2, respectively.  Panel (e) strongly resembles panel a); both depict Pathway A1 with weak core-envelope coupling.  The binaries in panel (e) are all circulating in the limit $f_{\rm B} \to 0$ where neither is spinning, but approach morphology fractions of about 70\% circulating and 30\% librating about $\pi$ for $f_{\rm B} \gtrsim 0.05$ as in panel (a).  Pathway B2 differs from A1 because of the higher $q_{\rm ZAMS}$, larger natal kicks, and accretion onto the primary WR star that gives it a maximal spin even for $f_{\rm B} = 0$.  It is important to note that in panel (f), unlike in panel (b), it is assumed that the Kerr limit can be preserved without mass loss, preventing MRR and yielding a much smaller BBH mass ratio $q \approx 0.63$.  As $f_{\rm B}$ increases, the secondary spin increases and the spin asymmetry of the binary is reduced.  The circulating fraction approaches 90\% consistent with binaries clustered in the upper right corner of the ($\cos\theta_{1\infty} - \cos\theta_{2\infty}$) plane for these parameters (see the top center panel of Fig.~14 of \cite{Gerosa2015b}).

These morphologies are generalizations of the spin-orbit resonances \cite{Schnittman2004}. Although prior dependent, some evidence of a preference for libration about $\pm\pi$ was found in the BBH population \cite{Varma2022b}. If confirmed, this suggests that some binaries originated from isolation with inherited spin or high primary BH spin from accretion in Pathway B2 before the onset of MRR, i.e.  $f_{\rm a} < f_{\rm a,MRR}$. Also, this could imply formation in Pathway A1 without MRR if an inherited primary spin and efficient tidal alignment of the secondary WR star spin create a large asymmetry between the misalignments \cite{Gerosa2013}, as shown in panel (e) of Fig.~\ref{F:Morphs} for nonzero $f_{\rm B}$ where the librating about $\pm\pi$ morphology is sub-dominant.

\subsection{Asymptotic Misalignments and Nutation}
\label{subsec:ResultsAsymp}

\begin{figure*}[!t] 
  \centering
  \includegraphics[width=1.0\linewidth]{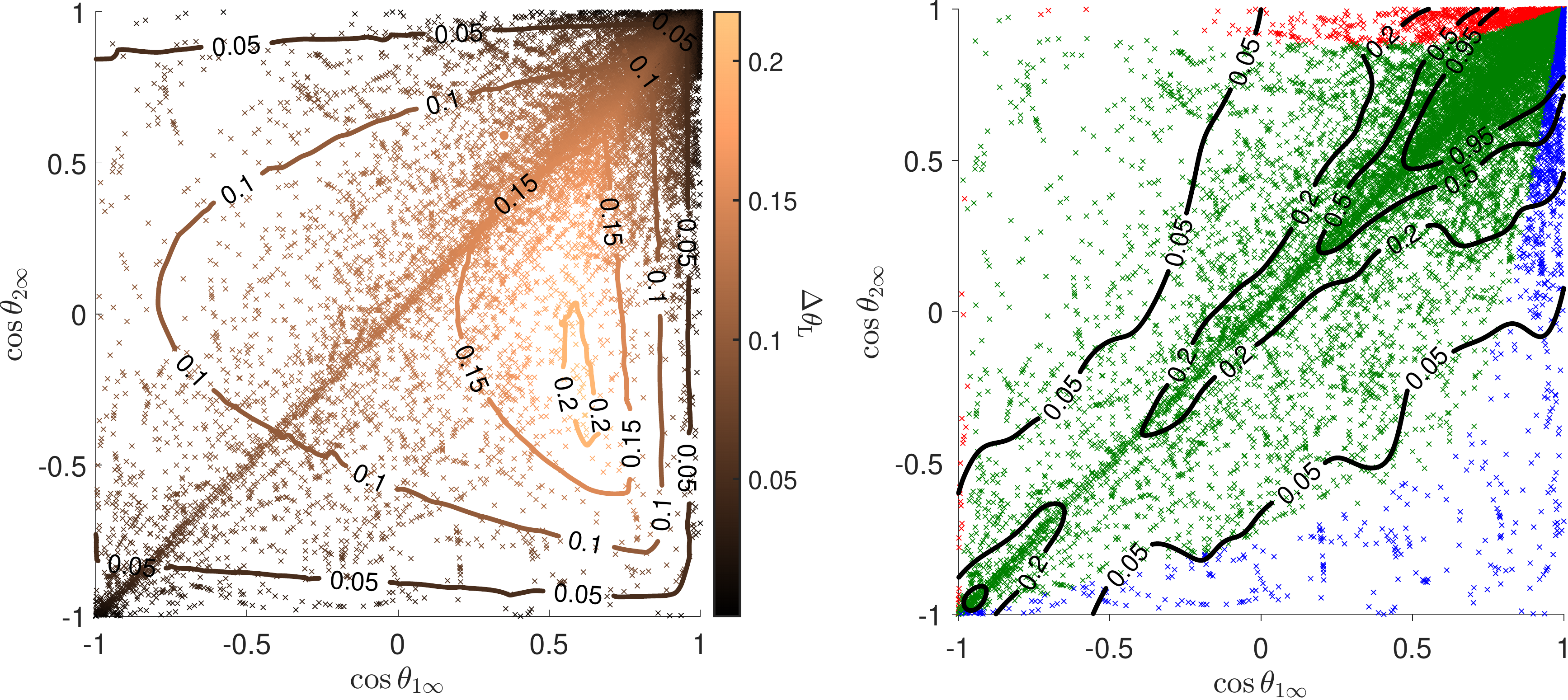}
   \caption{The plane composed of the asymptotic values of the spin-orbit misalignment cosines $\cos\theta_{1\infty}-\cos\theta_{2\infty}$, i.e. their values as $r/M \to \infty$, for BBHs that originate from Pathway B2, i.e., CEE1-SMT2-SN1-SN2, assuming weak core-envelope coupling with $f_{\rm B} = 0.1$ and negligible mass loss in BH formation due to the Kerr limit. The stellar binaries are initialized with $Z = 0.0002$, $a_{\rm ZAMS} = 15{,}000$~R$_{\odot}$, $m_{1,\rm ZAMS} = 70$~M$_{\odot}$, $m_{2,\rm ZAMS} = 60$~M$_{\odot}$, $f_{\rm a} = 0.2$, and $\sigma = 200$~km/s. The color of the data points in the left (right) panel corresponds to the value of the nutation amplitude $\Delta\theta_L$ (spin-precession morphology) at the end of the inspiral, i.e. $r_f/M \approx 16$ given by Eq.~(\ref{E:FinalSep}), where the BBH mass ratio is $q \approx 0.63$ and the spin magnitudes are maximal. The contour lines in the left (right) panel correspond to the value of $\Delta\theta_L$ (to the probability density).
   } \label{F:AsymptoticAngles}
\end{figure*}

The five precession parameters examined in subsection~\ref{subsec:ResultsParams} are a geometrically intuitive set of parameters that encode the rich variety of relativistic spin precession phenomenology.  They are constant on the precession timescale while evolving on the radiation-reaction timescale.  We evaluated the five precession parameters near the end of the BBH inspiral (i.e., see Eq.~(\ref{E:FinalSep})), where their signature on the emitted gravitational waveform is most pronounced. It can be illuminating, however, to also examine the spin-orbit misalignments in the limit of very large binary separation, i.e., $\cos\theta_1 \to \cos\theta_{1\infty}$ and $\cos\theta_2 \to \cos\theta_{2\infty}$ as $r/M \to \infty$. These asymptotic parameters are constants for each binary that determine the five precession parameters and precession morphology at all separations in the multi-timescale approximation. In this section, we approximate $\cos\theta_{1\infty}$ and $\cos\theta_{2\infty}$ as being equal to the values of $\cos\theta_{1}$ and $\cos\theta_{2}$ at the beginning (end) of the BBH inspiral (formation), which is reasonable to $\lesssim$ 1 part in 1000.

Figure~\ref{F:AsymptoticAngles} depicts the $\cos\theta_{1\infty} - \cos\theta_{2\infty}$ plane for the distribution of BBHs that evolve from Pathway B2 assuming weak core-envelope coupling with $f_{\rm B} = 0.1$ and negligible mass loss in BH formation due to the Kerr limit.  The stellar progenitors were initialized with $Z = 0.0002$, $a_{\rm ZAMS} = 15{,}000$~R$_{\odot}$, $m_{1,\rm ZAMS} = 70$~M$_{\odot}$, $m_{2,\rm ZAMS} = 60$~M$_{\odot}$, $f_{\rm a} = 0.2$, and $\sigma = 200$~km/s. These are the same BBHs at the largest value of $f_{\rm B}$ shown in Fig.~\ref{F:VsBreakB}, where the BBH mass ratio is $q \approx 0.63$, the total mass is $M = m_1 + m_2 \approx 30 + 19 = 49$~M$_{\odot}$, and the dimensionless BH spin magnitudes are maximal. If we instead assume isotropic mass loss in BH formation due to the Kerr limit, the BBH mass ratio would be higher and $\Delta\theta_L$ would be greatly suppressed.

In the left panel of Fig.~\ref{F:AsymptoticAngles}, the value of $\Delta\theta_L$ at the end of the inspiral, i.e., $r_f/M \approx 16$ given by Eq.~(\ref{E:FinalSep}), is displayed by the color bar and contour lines. The contour corresponding to $\Delta\theta_L = 0.2$ contains the location of the maximum of $\Delta\theta_L$ for this distribution of binaries. The moderate $q$ and maximal spin magnitudes for this choice of initial parameters yields a maximum nutation amplitude $\Delta\theta_L > 0.2$ consistent with the global maximum shown in the upper left panel of Fig.~3 of \cite{GangardtSteinle2021}.

The colors in the right panel of Fig.~\ref{F:AsymptoticAngles} correspond to the spin-precession morphology (red are librating about $\pi$, blue are librating about $0$, and green are circulating).  The solid lines are contours of constant probability density. BBHs are clustered in the top right corner because the ZAMS stellar spins are aligned with $\mathbf{L}$, and along the diagonal line from (-1, -1) to (1, 1) because the first natal kick gives the same misalignment to both binary components. The second natal kick provides the scatter off of the diagonal. The vast majority of BBHs in this distribution are in the circulating morphology, consistent with the bottom right panel of Fig.~\ref{F:Morphs} as $f_{\rm B} = 0.1$. 

These two sets of contour lines in the left and right panels of Fig.~\ref{F:AsymptoticAngles} show that a substantial subset of BBHs from the isolated channel can obtain significant nutation amplitude, i.e., $\Delta\theta_L \gtrsim 0.1$, most of which occupy the circulating morphology. There are few BBHs in the region of maximum $\Delta\theta_L$ near the point (0.5, -0.2).  It was previously noted that BBHs from the isolated channel are preferentially aligned ($\cos\theta_{i\infty} \approx 1)$ \cite{Gerosa2013,Gerosa2015b}, while those that evolve from the dynamical-formation channel have isotropic spins and are thus expected to uniformly fill the $\cos\theta_{1\infty} - \cos\theta_{2\infty}$ plane. Therefore, an observed population of BBHs from GW measurements that preferentially avoid occupying the region of this plane where $\Delta\theta_L$ is theoretically largest would be indicative of isolated binary formation.  Smaller values of $f_{\rm B}$ would lead to smaller BH spins and change the location of the $\Delta\theta_L$ peak, e.g., see Figs.~3 and 4 of \cite{GangardtSteinle2021}.

\section{Discussion} \label{sec:Discussion}

We have identified regions of the isolated binary parameter space from which BBHs that experience spin precession may emerge using five precession parameters that geometrically encapsulate spin-precession phenomenology.  Natal kicks with sufficiently large dispersion $\sigma$ allow for precession in each pathway, and the three mechanisms that provide high spin magnitudes determine whether precession is regular (without nutation) or is generic (with nutation). Parameters that measure precession,
i.e., $\langle\theta_L\rangle$ and $\langle\Omega_L\rangle$, can behave similarly for distributions of regular and generic binaries, but the parameters that measure nutation, $\Delta\theta_L,~\omega,~\Delta\Omega_L$, behave differently. The main conclusions from this study are as follows:
\begin{enumerate}
    \item Under the conservative assumption that stellar binary spins are initially aligned with their orbital angular momentum, natal kicks with dispersion $\sigma \gtrsim 30$~km/s (200~km/s) that occur before (after) CEE are needed to sufficiently misalign spins to produce significant precession.

    \item Weak stellar core-envelope spin coupling allows large natal spins (WR breakup spin fraction $f_{\rm B} \gtrsim 0.05$) which generically yield nutating systems.
    
    \item Accretion by the primary star as a BH (accreted fraction $f_{\rm a} \gtrsim 0.1$) or as a WR star ($f_{\rm a} \gtrsim 0.01$) produces highly precessing systems.
    
    \item Tides can produce large $\chi_{\rm eff}$ but generally do not provide large precession $\langle\theta_L\rangle$ or nutation $\Delta\theta_L$ amplitudes.  
    
    \item A large fraction of binaries can exhibit librating morphologies depending on the complications of phenomena such as accretion, mass-ratio reversal, and tidal synchronization and alignment.
    
    \item Binaries of isolated origin preferentially cluster in the upper right corner and along the diagonal of the $(\cos\theta_{1\infty}-\cos\theta_{2\infty})$ plane compared to those of dynamical origin, but can still obtain significant nutation amplitude $\Delta\theta_L$.
\end{enumerate}

It was traditionally thought that a measurement of nutation in GW data analysis would be a ``smoking gun'' for the dynamical formation channel of BBH origin \cite{Farr2017,Mapelli2020,Gerosa2021}; however, we find that significant nutation amplitude $\Delta\theta_{\rm L}$ is a key signature of BBHs that evolve from the isolated channel with weakly coupled stellar progenitors. This holds for all four evolutionary pathways we explore, implying that a measurement of nutation in GW data analysis might suggest isolated channel origin with weak core-envelope coupling and $f_{\rm B} \gtrsim 0.05$. Although the likely range of $f_{\rm B}$ is model dependent due to the uncertainties of stellar evolution, measurements of nutation from GW observations might constrain the likely range of $f_{\rm B}$. We do not vary the initial primary mass in the results of section~\ref{sec:Results} as was done in \cite{Steinle2021}, but our results presented here would be qualitatively similar for varied initial primary mass. In Scenario A, sufficient accretion onto the secondary star can increase the BBH mass ratio and suppress $\Delta\theta_{\rm L}$, which is more pronounced in Pathway A2 where the BBH mass ratio is larger. We suggest that parameter estimation of populations of BBHs, e.g., via hierarchical Bayesian inference, ought to include the possibility for inheritance of high BH spins.

The complicated dependence of $\chi_{\rm eff}$, $\langle \theta_L \rangle$, and $\langle\Omega_L\rangle$ on $f_{\rm a}$ must be taken into account when interpreting BBH formation using these precession parameters. The timescale over which SMT occurs is difficult to calculate. We assume that accretion is super-Eddington resulting in a high spin magnitude, consistent with recent population-synthesis modeling \cite{Zevin2022}. Assuming Eddington-limited accretion instead would suppress the BH spin, i.e., $\chi_{\rm BH} \lesssim 0.1$ \cite{Zevin2022}, and the precession amplitude $\langle \theta_L \rangle$ in Pathway B1, however this may be more complicated in Pathway B2 as the Eddington limit of WR stars is uncertain \cite{Maeder2012b}.

We treat the natal kick strength $\sigma$ as a free parameter and assume that $\sigma$ is the same in the natal kicks of the primary and secondary stars. Instead, if $\sigma$ differed for the two natal kicks, then the average BBH may more generically be born with differential misalignments, as is possible in A1 due to tidal alignment with the same value of $\sigma$ in each natal kick, allowing for interesting signatures of spin precession \cite{Schnittman2004,Kesden2015,Gerosa2015b}. 
Observationally, $\sigma$ is not well constrained for black holes, e.g., see \cite{Callister2021,Stevenson2022}. A recent claim that a microlensing event has uncovered the first unambiguous detection of an isolated black hole suggests that black holes may receive natal kicks. The black hole has a mass $\approx 7$ M$_{\odot}$, and its measured space velocity implies that $\sigma \approx 45$ km/s if it formed as the result of stellar evolution \cite{Sahu2022}. Subsequent analyses have shown that $\sigma \lesssim 100$ km/s \cite{Andrews2022} and such a system likely originated from a binary \cite{Vigna2022}. In our model, such an isolated black hole could emerge from a binary system that evolves through Pathway A1 after being unbound by a natal kick, or from a binary whose secondary Wolf-Rayet star was destroyed from Roche-lobe overflow after CEE.

If black holes do not receive natal kicks, then another source of misalignments will be needed to obtain precessing isolated BBHs. One observationally uncertain possibility is that the ZAMS binary star forms with random spin directions. These would likely be retained until BBH formation if alignment from tides or accretion is avoided \cite{Postnov2018}.   They would result in generic precession \cite{GangardtSteinle2021} which would be difficult to distinguish from the spin precession of BBHs of dynamical origin. Another uncertain possibility might exist from SMT, assuming strong core-envelope coupling and $\gtrsim 30\%$ of the mass of the donor is ejected, where the donor star's spin direction can nearly "flip" into the orbital plane independent of its prior misalignment \cite{Stegmann2021}. In the context of our model, this implies that BBHs from Scenario A, where the primary star initiates SMT, or Scenario B, where the secondary star initiates SMT, would both exhibit regular precession even in the absence of natal kicks. If the amount of mass retained is large, accretion can easily cause mass-ratio reversal in pathways A1 and B2 leading to interesting correlations between the masses and spin directions. Yet another possibility is that the BH spins themselves may be "tossed" in core-collapse formation although the mechanism that provides such a torque is currently not clear \cite{Tauris2022}.

The results of this work constitute the first systematic study of the dependence of the spin precession of isolated BBHs itself, rather than just misaligned spins,
on various astrophysical uncertainties and assumptions. We hope that this study can help to motivate further investigation into the possibility of observing precession and nutation in GW data of BBHs, which may uncover the likely formation channels of the BBH population. An evolutionary scenario where both binary components lose their envelopes through stable mass transfer \cite{Broekgaarden2022} could result in a nutating BBH if efficient accretion yields two high spins. Also, the chemically homogeneous evolution scenario may provide nutating BBHs if the tidally or rotationally induced high spins are retained \cite{Marchant2016,Cui2018}. A future follow-up of our analysis using a population-synthesis approach could further explore our conclusions. Although we do not explore it here, the effective precession parameter $\chi_{\rm p}$ \cite{Schmidt2015} is often used to study the spin precession of BBHs. The generalized form of this parameter \cite{Gerosa2020} is analogous to the precession amplitude $\langle\theta_L\rangle$, implying that it would behave similarly as a function of the initial stellar parameters.

Constraints on the five precession and nutation parameters of individual LIGO/Virgo sources demonstrate that precession is indeed present for certain systems, but that nutation has yet to be detected \cite{Gangardt2022}. A work in preparation will explore Bayesian parameter estimation of these parameters and compute realistic uncertainties \cite{Stoikos2022}.

\begin{acknowledgments}
The authors would like to thank Davide Gerosa for helpful discussion and the referees for insightful comments. N.S. is supported by Leverhulme Trust Grant No. RPG-2019-350, European Union's H2020 ERC Starting Grant No. 945155-GWmining, and Cariplo Foundation Grant No. 2021-0555, and both NS and MK were supported by the National Science Foundation Grant No. PHY-1607031.
\end{acknowledgments}

\bibliography{bibme}{}

\begin{thebibliography}{59}
\bibitem [{\citenamefont {{Stoikos}}\ and\ \citenamefont
  {{Kesden}}()}]{Stoikos2022}%
  \BibitemOpen
  \bibfield  {author} {\bibinfo {author} {\bibfnamefont {E.}~\bibnamefont
  {{Stoikos}}}\ and\ \bibinfo {author} {\bibfnamefont {M.}~\bibnamefont
  {{Kesden}}},\ }\href@noop {} {\bibinfo {title} {{Parameter estimation of spin
  precession parameters for black hole binaries (to be
  published)}}}\BibitemShut {NoStop}%
\end{thebibliography}

\end{document}